\documentclass[11pt]{article}
\pdfoutput=1
\usepackage{dcolumn}% Align table columns on decimal point
\usepackage{bm}% bold math
\usepackage{graphicx}% Include figure files
\usepackage[numbers,sort&compress]{natbib}    
\usepackage{amssymb,amsmath,mathrsfs}
\usepackage{multirow}
\usepackage{subfig}
\usepackage{graphbox}
\usepackage{color,url}
\usepackage[font=footnotesize,labelfont=bf,justification=justified]{caption}
\usepackage[colorlinks=true
,urlcolor=blue
,anchorcolor=blue
,citecolor=blue
,filecolor=blue
,linkcolor=blue
,menucolor=blue
,linktocpage=true
,pdfproducer=medialab
,pdfa=true
]{hyperref}
\usepackage{slashed}
\usepackage{rotfloat}

\allowdisplaybreaks

\begin{document}
\begin{flushright}
FTUV-19-0620.6612 \\
IFIC/19-30
\end{flushright}
\raggedbottom
\thispagestyle{empty}

\def\thefootnote{\fnsymbol{footnote}}

%\begin{flushright}
%IFT-UAM/CSIC-19-XX\\
%FTUAM-19-YY\\
%LPT-Orsay-19-ZZ
%\end{flushright}

%\vspace*{2cm}
%
%\title{{\boldmath LHC sensitivity to singly-charged scalars decaying into electrons and muons}}
%
%\def\uv{\affiliation{\it Departament de F\'{\i}sica Te\`orica, Universitat de Val\`encia and IFIC, \\
%Universitat de Val\`encia-CSIC, Dr. Moliner 50, E-46100 Burjassot (Val\`encia), Spain \vspace*{5mm}}}
%
%\def\laplata{\affiliation{\it IFLP, CONICET - Dpto. de F\'{\i}sica, Universidad Nacional de La Plata, \\ 
%C.C. 67, 1900 La Plata, Argentina \vspace*{2mm}}}
%
%\author{\sc Julien~Alcaide}
%\email{julien.alcaide@uv.es}
%\uv
%
%\author{\sc Nicol\'as~I.~Mileo \vspace*{4mm}}
%\email{mileo@fisica.unlp.edu.ar}
%\laplata

\begin{center}
\vspace*{8mm}
\begin{Large}
\textbf{\textsc{LHC sensitivity to singly-charged scalars decaying into electrons and muons}}
\end{Large}

\vspace{1cm}

{\sc
Julien~Alcaide$^{1}$%
\footnote{{\tt \href{mailto:julien.alcaide@uv.es}{julien.alcaide@uv.es}}}%
and Nicol\'as~I.~Mileo$^{2}$%
\footnote{{\tt \href{mailto:mileo@fisica.unlp.edu.ar}{mileo@fisica.unlp.edu.ar}}}%
}

\vspace*{.7cm}

{\sl

$^1$Departament de F\'{\i}sica T\`eorica, Universitat de Val\`encia and IFIC, \\
Universitat de Val\`encia-CSIC, Dr. Moliner 50, E-46100 Burjassot (Valencia), Spain

\vspace*{0.1cm}

$^2$IFLP, CONICET - Dpto. de F\'{\i}sica, Universidad Nacional de La Plata, \\ 
C.C. 67, 1900 La Plata, Argentina
}

\end{center}

\vspace{0.1cm}

\begin{abstract}
\noindent
Current LHC searches for non-supersymmetric singly-charged scalars, based on Two-Higgs-Doublet models, in general focus the analysis on third-generation fermions in the final state. 
However, singly-charged scalars in alternative extensions of the scalar sector involve Yukawa couplings not proportional to the mass of the fermions. Assuming the scalar decays into electrons~and~muons, it can manifest cleaner experimental signatures.
In~this~paper we suggest that a singly-charged scalar singlet, with electroweak production, can start to be probed in the near future with dedicated search strategies. 
Depending on the strength of the Yukawa couplings, two independent scenarios are 
considered: direct pair-production
% through weak gauge bosons 
 (small~couplings) 
and single-production via virtual neutrino exchange
(large~couplings).
We show that, up to a mass as large as 500~GeV, most of the parameter space could be excluded at the 95\%~C.L. in a high-luminosity phase of the LHC.
Our results also apply to other frameworks, provided the singly-charged scalar exhibits similar production patterns and dominant decay modes. 
\end{abstract}
%\maketitle
\thispagestyle{empty}
%\maketitle
\def\thefootnote{\arabic{footnote}}
\setcounter{page}{0}
\setcounter{footnote}{0}
%\maketitle
\newpage

\section{Introduction}
\label{intro}
%\ja{Parrafo del SM, enormous success, incomplete, extensions, singlet bla bla bla}
A clear evidence of physics beyond the Standard Model (SM) would be doubtlessly the existence of charged scalars, as predicted in many extensions of the SM that incorporate, for instance, new weak multiplets to the scalar sector. The charged Higgs present in the Two Higgs Doublet Model (2HDM) provides a widely studied example of a singly-charged scalar arising from an additional weak doublet~\cite{Branco}. In fact, most of the searches at colliders are based on the 2HDM
and in general focus on third-generation fermions in the final state. The charged Higgs boson has been searched at the LEP considering final states involving the decay channels $H^+\to c\bar{s}$ and $H^+\to \tau^+\nu_{\tau}$, and assuming that these saturate the decay branching ratio. The combination of all the LEP data excludes a charged Higgs with mass below 80~GeV within the context of the 2HDM~\cite{LEP}. On the other hand, the searches for the charged Higgs at the LHC target on masses above the LEP limit and typically consider that it is produced in top-quark decays or in association with a top quark and/or a bottom quark. These searches cover not only the decay channels analized at the LEP~\cite{cs,taunu}, but also the mode $H^+\to t\bar{b}$, when the charged Higgs is heavier than the top quark~\cite{tb1,tb2}.\par
Charged scalars also appear in models that incorporate weak triplets such as the Georgi-Machacek model~\cite{Georgi}, in which the scalar sector of the SM is extended with	 one complex and one real weak triplets giving rise to both singly and doubly-charged scalars, or the Type-II seesaw models~\cite{seesawII1,seesawII2,seesawII3,seesawII4,seesawII5}, where only one weak triplet is added whose neutral component acquires a vacuum expectation value and generates a Majorana mass for the neutrinos. The mass and couplings of the singly-charged scalar appearing in the Georgi-Machacek model have been tested at the LHC through the search for resonant $WZ$ production in the fully leptonic final state~\cite{resonanceWZ} and also in the decay channel $H^+\to W^{+}Z$, with the charged Higgs produced via vector boson fusion~\cite{WZ}. The doubly-charged scalar has been, in turn, searched for in the decay channel $H^{++}\to W^+W^+$~\cite{WW}.\par  
Finally, weak-singlet singly-charged scalars are particularly interesting since they appear in models that generate small radiative neutrino masses. Prominent examples of these models are the Zee model~\cite{ref:Zee_model} and the Zee-Babu model~\cite{ref:Zee-Babu_model1,ref:Zee-Babu_model2} in which the neutrino masses are generated at one and two loops, respectively. In addition, the $SU(2)_{L}\times U(1)_Y$ gauge quantum numbers of the charged scalar weak singlet are the same as those of the right-handed sleptons in supersymmetric models~\cite{Martin} and the singly-charged scalar in $L-R$ symmetric models~\cite{ref:L-Rmodel}. So far, there are no dedicated searches for the weak-singlet singly-charged scalar at the LHC. With this motivation, in this paper we develop a LHC search strategy to set bounds on the mass of the charged scalar weak singlet, focused in processes with leptons in the final state, since in principle these are more promising than those involving quark decay modes due to the impact of the large QCD backgrounds. Although throughout the text we will concentrate on the weak-singlet singly-charged scalar, our results can be easily extended to other models of charged scalars with similar production and decay patterns.\par
%In this paper we suggest a LHC search strategy to set bounds on the mass of the charged scalar, focused in processes with leptons in the final state.
In general, the leptonic interaction of the charged scalar, called from now on $h^\pm$, would be with pairs of leptons with either chirality. Without adding right-handed neutrinos to the field content\footnote{If we considered right-handed neutrinos there would be two additional operators, namely
\begin{equation*}
\mathcal{O}_{RR} = \bar{\nu}_{Ra}e_{Rb}^c h^- + \mathrm{h.c.} + ...
\end{equation*}
\begin{equation*}
\mathcal{O}_{RL} = \bar{\nu}_{Ra}e_{Lb} h^+ + \mathrm{h.c.} + ...,
\end{equation*}
whose implementation would be equivalent to that of operators LL and LR, respectively. 

%En ambos casos tiene que haber un helicity flip del $\bar\nu_L$ que sale del acoplamiento con el W en un $\bar\nu_R$, a traves de una insercion de masa de Dirac.
}, two different combinations of chiralities arise:
\begin{equation}
\label{eq:LLlagrangian}
\mathcal{O}_{LL} = \bar{\nu}_{La}e_{Lb}^c h^- + \mathrm{h.c.} + ...
\end{equation}
\begin{equation}
\label{eq:LRlagrangian}
\mathcal{O}_{LR} =	  \bar{\nu}_{La}   e_{Rb} h^+ + \mathrm{h.c.} + ...,
\end{equation}
where the subindex $L$ ($R$) stands for the left-handed (right-handed) component of the field, and the superindex $c$ transforms the given field into its charge conjugate counterpart. In the context of renormalizable theories, the LL operator comes from gauge invariant leptonic interactions of either a weak singlet or a triplet, while the LR operator is realised when the interaction is with a doublet. Furthermore, both operators can also be generated within effective field theories (EFT) in which the heavy states have been integrated out.\par
In the specific case of the singlet, the renormalizable Lagrangian describing its Yukawa interactions with leptons is given by:
%Although all our conclusions can be extend for models with similar production patterns, throughout the text we will focus the analysis around a charged scalar \textit{weak singlet}. The $SU(2)_L\times U(1)_Y$ quantum numbers of such scalar are the same as those of the right-handed sleptons in supersymmetric models [...], the singly-charged scalar in the Zee model \cite{ref:Zee_model} as well as in the Zee-Babu model \cite{ref:Zee-Babu_model1,ref:Zee-Babu_model2} and in L-R symmetry models \cite{ref:L-Rmodel} \ja{mi amiga que hace L-R models me dijo que solo encontraron un paper anterior al de ellas en el que usan singletes simplemente cargados, pero que creen que esta mal}.
%The renormalizable Lagrangian describing the Yukawa interaction of the singlet to leptons is given by:
\begin{equation}
\label{eqLag}
%\mathcal{L}_{h^{\pm}}=(D_{\mu}h^{+})^{\dagger}(D^{\mu}h^{+})-m^2_{h^{+}}|h^{+}|^2-\frac{\lambda_h}{2}|h^{+\dagger}h^{+}|^2-\lambda_{hH}h^{+\dagger}h^{+}H^{\dagger}H+(f_{ab}\bar{\ell}_{La}\ell^{\,c}_{Lb}h^+ + \mathrm{h.c.}),
\mathcal{L}_{h^{\pm}}^{LL} = f_{ab}\bar{\ell}_{La}\ell^{\,c}_{Lb}h^+ + \mathrm{h.c.},
\end{equation}
where $\ell_L$ denotes the lepton doublet and $\ell_L^c=i\sigma_2 \ell_L^*$ its charge conjugate field. The Yukawa coupling $f_{ab}$ (with $a,b=e,\mu,\tau$) needs to be antisymmetric due to the presence of the Clebsh factor $i\sigma_2$. This interaction belongs to the LL category. 

Alternatively, the LR interaction of the singlet to leptons can be realised at dimension 5 following the effective Lagrangian:
\begin{equation}	
\label{eq:LagLR}
%\mathcal{L}_{h^{\pm}}=(D_{\mu}h^{+})^{\dagger}(D^{\mu}h^{+})-m^2_{h^{+}}|h^{+}|^2-\frac{\lambda_h}{2}|h^{+\dagger}h^{+}|^2-\lambda_{hH}h^{+\dagger}h^{+}H^{\dagger}H+(f_{ab}\bar{\ell}_{La}\ell^{\,c}_{Lb}h^+ + \mathrm{h.c.}),
\mathcal{L}_{h^{\pm}}^{LR} = {c_{ab}\over\Lambda} \left(\bar{\ell}_{La}\tilde H e_{Rb}h^+\right) + \mathrm{h.c.},
\end{equation}
where $\tilde H = i\sigma_2 H^*$, $c_{ab}$ encodes the information on new degrees of freedom, and $\Lambda$ is the scale of new physics. After spontaneous symmetry breaking, we retrieve the operator LR in Eq.~(\ref{eq:LRlagrangian}) and identify the coupling constant as $g_{ab}\equiv c_{ab} {v\over\Lambda}$, where $v\sim174$~GeV is the Higgs vacuum expectation value fixing the electroweak scale. Even for a $c_{ab}$ of order 1, a new physics scale around the TeV will typically introduce a suppression factor.  

We point out that, at leading order, a charged scalar interacting with leptons could be singly-produced (via Yukawa couplings) as well as pair-produced (via gauge interactions). In particular, a channel with two charged leptons in the final state ($2\ell$~channel) will involve both production modes, and their relative contribution depending on the strength of the Yukawa couplings. 
For sufficiently small values, the contribution of the single production can be neglected, and in fact this is the approach followed in Ref.~\cite{Cao:2017ffm}. As the Yukawa couplings grow, however, both contributions start to compete which makes troublesome to obtain conclusions from a search strategy based on this final state. 
%However, this strategy is best oriented to probe pure production channels and thus it is troublesome to obtain conclusions in a framework of large couplings. 
Therefore, we also analyse the three charged lepton channel ($3\ell$~channel) which is more suitable to explore the large Yukawa coupling scenario since it involves only the single production mechanism, via virtual neutrino exchange.\par 
Even when there are no dedicated searches of the singly-charged weak singlet at the LHC, the LEP and LHC searches for charged Higgs bosons and sleptons could in principle provide bounds. The LEP limits on a charged Higgs assume that $h^+$ decays exclusively to $c\bar{s}$ and $\tau^+\nu_{\tau}$ and in the case in which $h^+$ can decay significantly into electrons and muons it is shown in Ref.~\cite{Cao:2017ffm} that the combination of these limits with the slepton search results only exclude masses below 65~GeV. On the other hand, the current LHC searches for charged Higgs bosons consider production mechanisms and decay modes involving only third generation fermions, and then do not impose relevant constraints on a charged scalar decaying mostly into electrons and muons. This conclusion is even more radical if we assume that the charged scalar does not interact with quarks. Regarding the slepton searches at the LHC, those performed in the dilepton mode ($e^+e^-$ and $\mu^+\mu^-$) impose bounds in the plane $m_{\tilde{\chi}^0}$-$m_{\tilde{\ell}}$ for right-handed, left-handed and both right- and left-handed selectron and smuon production. From these bounds, only those corresponding to a massless neutralino and right-handed sleptons could be applied to our case. In Ref.~\cite{dilepton2018}, for example, masses above $\sim$ 380~GeV are excluded. However, this bound is obtained by assuming degenerate selectrons and smuons, while in our case the dilepton mode originates from a single charged scalar, which reduces the cross section of the process considerably and then weakens the limit. We note that these slepton searches do not consider the virtual lepton exchange production mode. Finally, the LHC searches of staus in the ditau mode can also impose constraints on the branching ratio of the singly-charged scalar into taus. Even when the current limits are not very restrictive~\cite{ditau2019}, they are very useful in order to cover the region of small decay branching ratio of $h^{\pm}$ into electrons and muons, where the search strategies based on the dilepton channel lose sensitivity. However, in this paper we will consider only electrons and muons in the final state and leave the analysis of topologies involving tau leptons as future work.\par 
%Hablar de busquedas de sleptones y lo que vimos nosotros de que no aplican xq suman left y right inclusivo. 

Our paper will be laid out as follows. In Section~\ref{sec:motivation} we discuss the production patterns of the charged scalar as well as its decay modes. In Section~\ref{sec:2lep} we describe the strategy performed in the $2\ell$-channel and set bounds on the mass of the scalar and its branching ratio to electrons and muons. In Section~\ref{sec:3lep}, we present the analysis in the $3\ell$-channel and sett limits on the Yukawa coupling and the mass of the scalar. Subsequently, we compare our findings using the LR operator with those obtained with the LL operator. Finally, we conclude in Section~\ref{sec:conclusions}.

\section{Production and dominant decay modes}
\label{sec:motivation}
The production of a charged scalar that does not interact with quarks will depend exclusively on its interactions with the gauge bosons and the leptons. The quantum numbers of the charged scalar under the SM gauge group completely determine the former while only set the tree-level structure of the latter but not its strength. In the case of the weak singlet, for example, the gauge symmetry enforces the renormalizable coupling $f_{ab}$ to be antisymmetric (see Eq.~(\ref{eqLag})). However, the strength of these couplings remains undetermined and must be constrained experimentally. For instance, since the interaction $\bar{\ell}_L\ell^c_L h^{+}$ induces charged lepton flavor violation, the couplings $f_{ab}$ will receive stringent constraints from lepton rare decays, which can be summarized in the requirements $|f_{e\mu}f_{\mu\tau}|,|f_{e\mu}f_{e\tau}|,|f_{e\tau}f_{\mu\tau}|\lesssim\mathcal{O}(10^{-5})~$\cite{McLaughlin:1999rr,Herrero-Garcia:2014hfa,TheMEG:2016wtm,Tanabashi:2018oca}. These bounds could be satisfied by considering that all the couplings $f_{ab}$ are highly suppressed and then that the renormalizable interactions involving them are negligible. In fact, this is the approach used in Ref.~\cite{Cao:2017ffm}, where both the interactions of the charged scalar with leptons and quarks are derived from dimension-5 operators. Another possibility would be to set two of the couplings to zero and left the remaining one unbounded. This scenario may be accomplished by imposing a lepton number global symmetry such as $L_i-L_j$, with $(i,j)=(e,\mu),(e,\tau)$ or $(\mu,\tau)$, which leads to $f_{ij}\neq 0$ and $f_{ab}=0$ for $ab\neq ij$. In this manner the above experimental bounds are satisfied with the dominant production and decay modes still driven by the renormalizable interactions of Eq.~(\ref{eqLag}). This will be the framework adopted in this paper, although the results can be applied in general to other models of singly-charged scalars with a similar pattern of production and decay channels.\par
\begin{figure}[H]
\begin{center}
\subfloat[]{\includegraphics[align=c,scale=0.54]{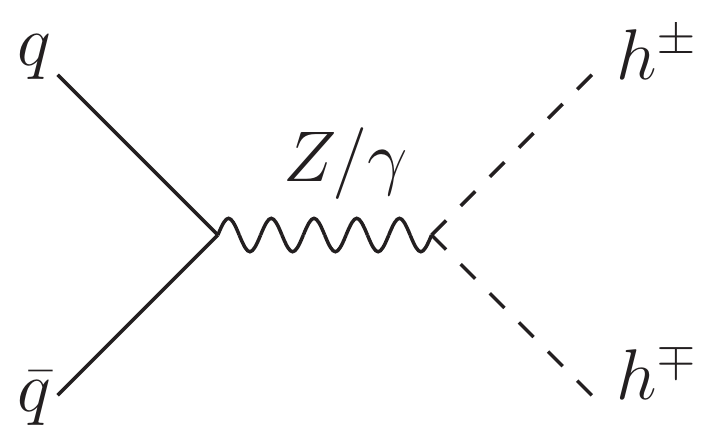}}\hspace*{1mm}
\subfloat[]{\includegraphics[align=c,scale=0.54]{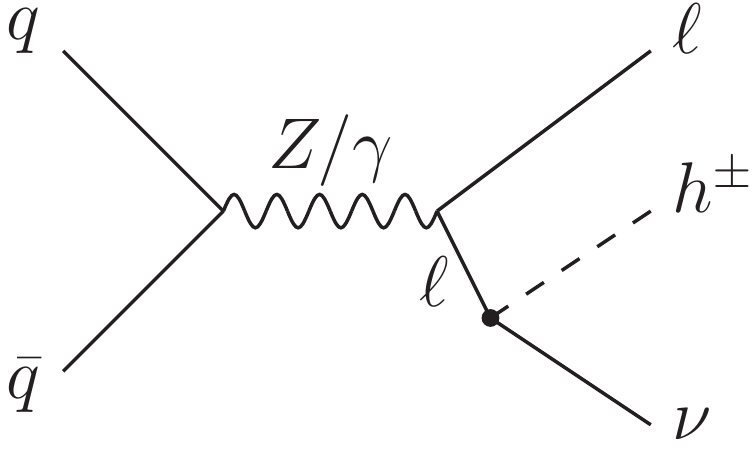}}\hspace*{1mm}
\subfloat[]{\includegraphics[align=c,scale=0.54]{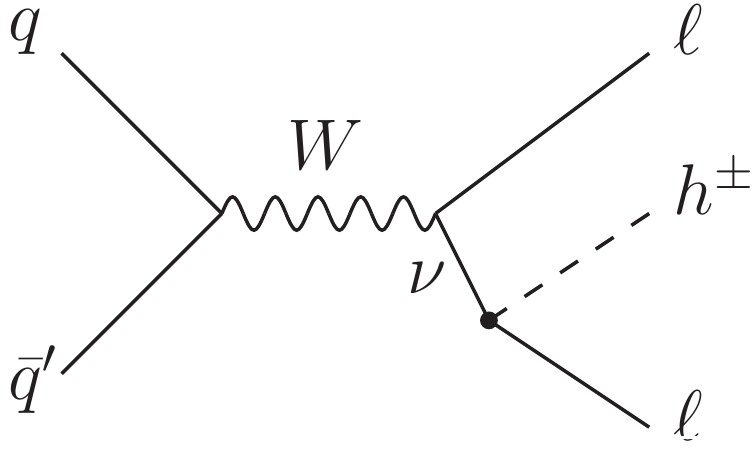}}\hspace*{1mm}
\subfloat[]{\includegraphics[align=c,scale=0.54]{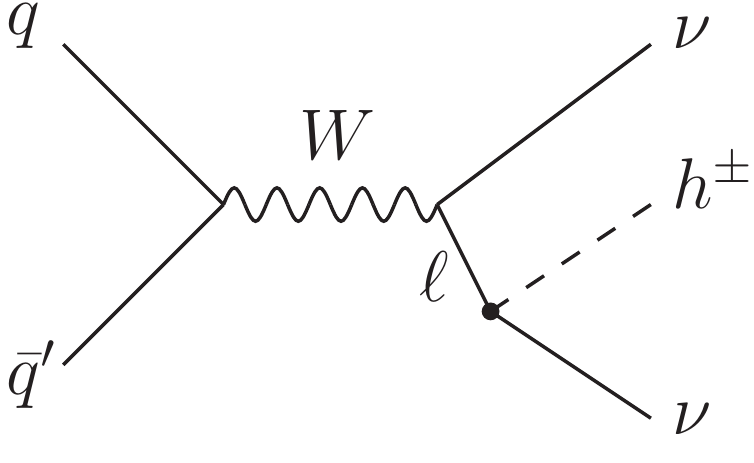}}
\caption{Feynman diagrams involved in the production of the singly-charged scalar: (a) corresponds to pair production, (b)-(d) depict different single production modes. The diagrams (a) and (b) contribute to the 2$\ell$+$E^{\mathrm{miss}}_T$ topology, while (c) and (d) lead to 3$\ell$+$E^{\mathrm{miss}}_T$ and 1$\ell$+$E^{\mathrm{miss}}_T$ topologies, respectively.}
\label{feymandiag}
\end{center}
\end{figure}
\begin{figure}[H]
\begin{center}
%\begin{tabular}{cc}
%\end{tabular}
\subfloat{\includegraphics[align=c,scale=0.54]{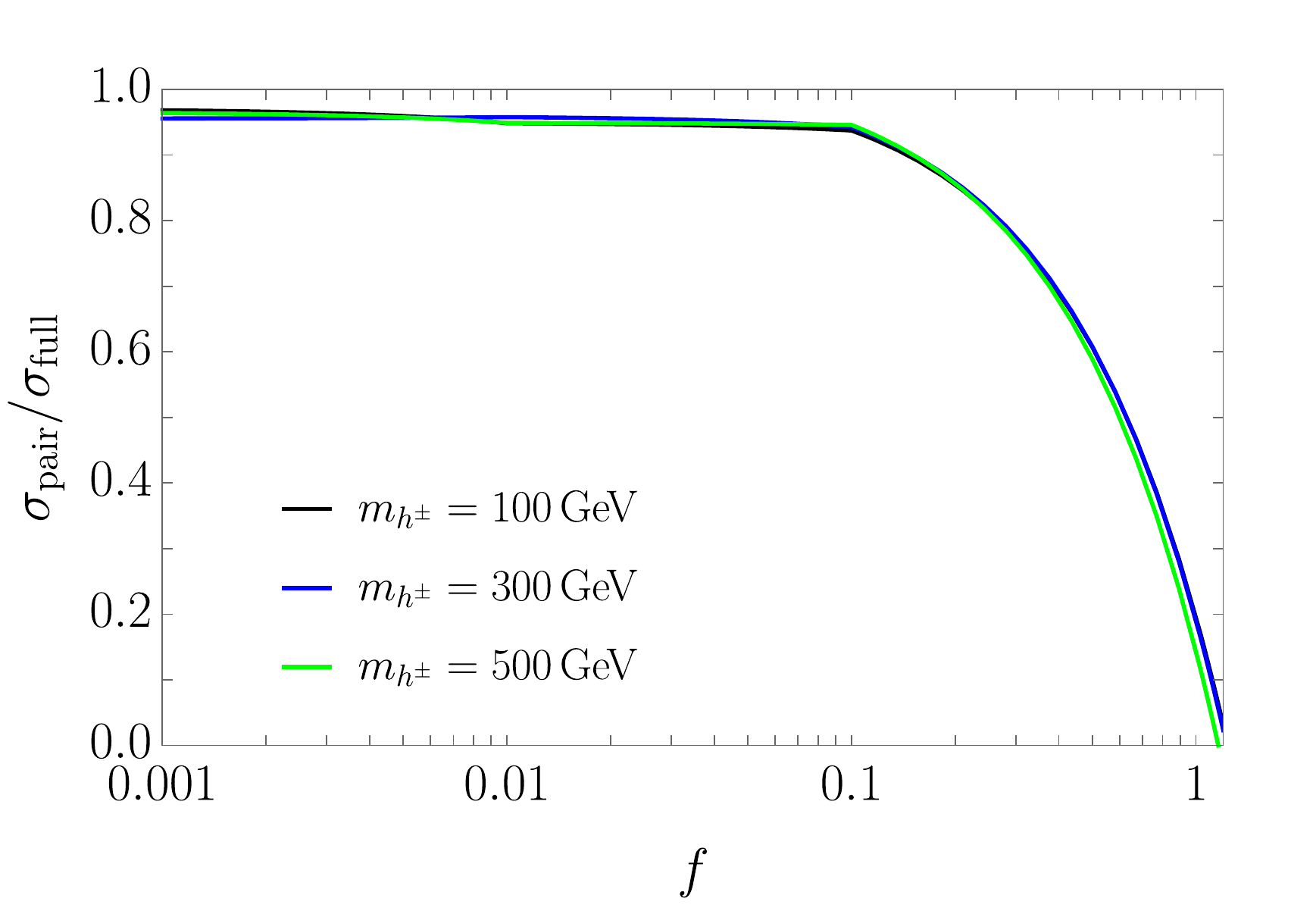}}\centering\subfloat{\includegraphics[align=c,scale=0.56]{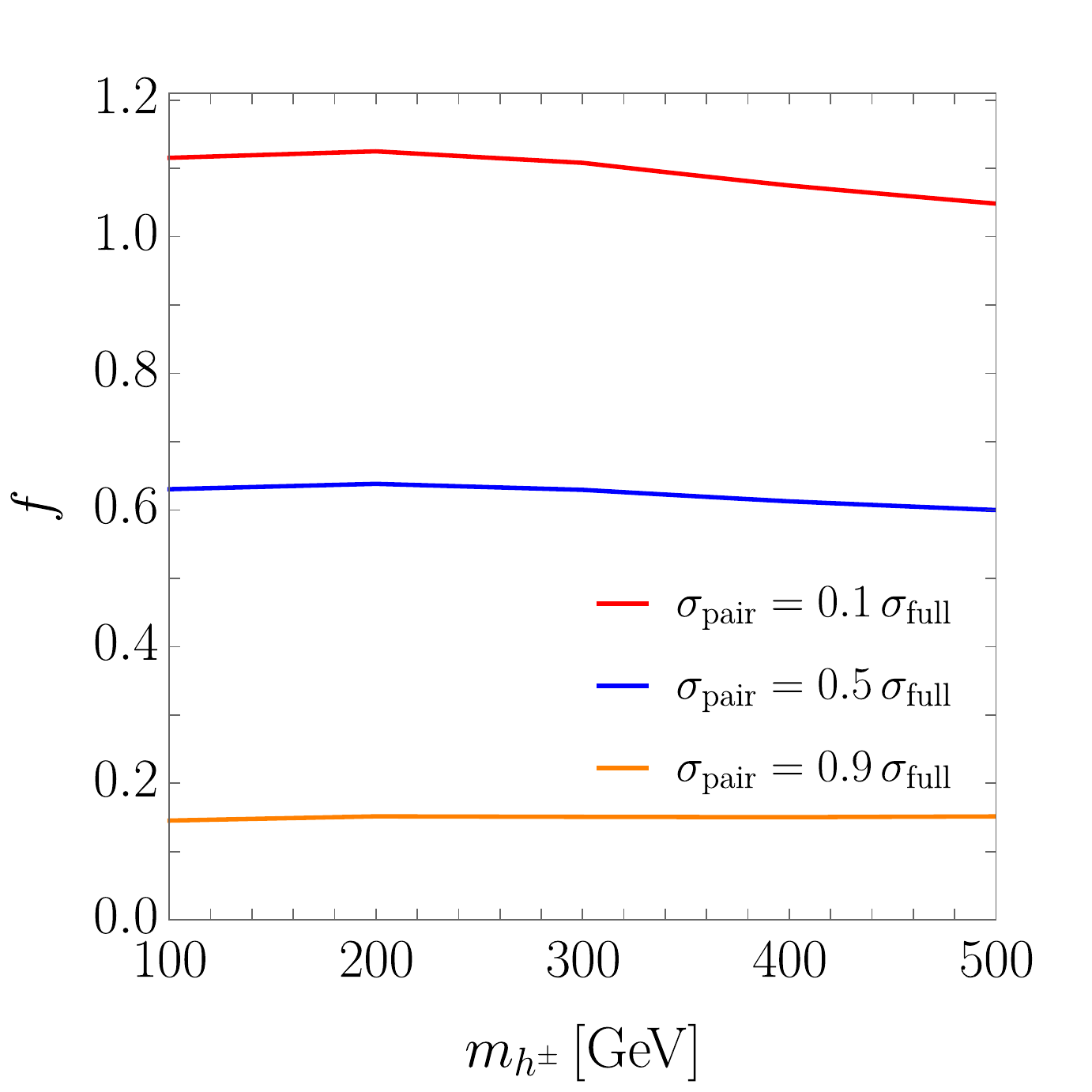}}
\caption{Contribution of the pair production mode ($\sigma_{\mathrm{pair}}$) to the cross section of the 2$\ell$ channel ($\sigma_{\mathrm{full}}$). Left panel: behavior of the ratio $\sigma_{\mathrm{pair}}/\sigma_{\mathrm{full}}$ with the coupling $f$ for three different masses of the charged scalar. Right panel: contours of $\sigma_{\mathrm{pair}}/\sigma_{\mathrm{full}}$ in the $f-m_{h^{\pm}}$ plane.}
\label{figXspairvsFull}
\end{center}
\end{figure}

The charged scalar can be produced in pairs through its interactions with the gauge bosons $Z/\gamma$ but can also be radiated from a lepton external leg in $s$-channel diagrams with $Z,\gamma
$ or $W$ bosons. This leads to the single production in association with two neutrinos, one neutrino and one charged lepton or two charged leptons. Unlike the pair production, the single production depends on the coupling of $h^{\pm}$ to leptons. The Feynman diagrams corresponding to the different production modes are shown in Figure~\ref{feymandiag}. Since we are assuming that the charged scalar does not couple to quarks, the dominant decay modes will be $h^{+}\to \ell^+ \nu_{\ell^{\prime}}$, with $\ell,\ell^{\prime}=e,\mu,\tau$. For the weak singlet with the renormalizable Lagrangian of Eq.~(\ref{eqLag}) we necessarily have $\ell\neq\ell^{\prime}$. Combining the possible decay channels with the production modes displayed in Figure~\ref{feymandiag}, three different final state arise:
\begin{itemize}
\item[1.-] Two opposite-sign leptons plus missing transverse energy, $2\ell + E^{\mathrm{miss}}_T$, associated to diagrams mediated by $Z/\gamma$ (diagrams (a) and (b)).
\item[2.-] Three leptons plus missing transverse energy, $3\ell + E^{\mathrm{miss}}_T$, associated to diagram (c).
\item[3.-] One lepton plus missing transverse energy, $1\ell + E^{\mathrm{miss}}_T$, associated to diagram (d). 
\end{itemize}
In this study we focus on the first two, which are in principle more promising than the third one that contains just one lepton in the final state. In addition, we consider only electrons and muons in the final state since these flavors can be treated inclusively given their similar cut efficiencies at the LHC. Final states involving tau leptons need to be treated separately and we left their study as future work. Thus, in the following, the term {\it lepton} will describe exclusively electrons and muons. 
%%%%
Since the 3$\ell$ channel originates exclusively from diagram (c) in Figure~\ref{feymandiag}, the corresponding cross section scales with the coupling of $h^{\pm}$ to leptons as: 
\begin{equation}
\label{eqscale3lep}
\sigma_{3\ell}(f) = f^2\times \sigma_{pp\to h^{\pm}\ell\ell}(f=1)\times \mathrm{BR}_{e+\mu},
\end{equation}
where $\sigma_{pp\to h^{\pm}\ell\ell}(f=1)$ is the production cross section for $f=1$, and $\mathrm{BR}_{e+\mu}\equiv\mathrm{BR}_{e}+\mathrm{BR}_{\mu}$ is the decay branching ratio of $h^{\pm}$ into electrons and muons. Clearly, a search strategy based on this channel would lose its sensitivity for decreasing values of the coupling~$f$. This is not the case for the $2\ell$ channel since, besides the production in association with $\ell \nu$, it also involves the pair production which in fact is the dominant mode for sufficiently small values of the coupling $f$. This can be established more precisely by considering the contribution of the pair production mode to the total cross section in terms of the coupling $f$ for different masses of the charged scalar. From the left panel of Figure~\ref{figXspairvsFull}, we see that the pair production mode gives the dominant contribution for $f$ below $0.1$. Above this value, the contribution of the single production starts to increase until becoming competitive with the pair production mode for $f\gtrsim 0.6$. The mild dependence with the charged scalar mass is due to the fact that the pair production cross section suffers a higher phase space suppression than the single production. This effect is more visible in the right panel of Figure~\ref{figXspairvsFull} where we plot various contours of the ratio $\sigma_{\mathrm{pair}}/\sigma_{\mathrm{full}}$ in the $f-m_{h^{\pm}}$ plane. \par
In summary, for $\mathcal{O}(f)\leqslant 0.1$ the cross section of the $2\ell$ channel is fully dominated by the pair production mode and then can be written as
\begin{equation}
\label{eqscale2lep}
\sigma_{2\ell} = \sigma_{pp\to h^{+}h^{-}}\times \mathrm{BR}^2_{e+\mu},
\end{equation}
while for $\mathcal{O}(f)= 1$ the scaling is the one given in Eq.~(\ref{eqscale3lep}) but with $\sigma_{pp\to h^{\pm}\ell\ell}$ replaced by $\sigma_{pp\to h^{\pm}\ell\nu}$. Taking the above discussion into account, we decided to separate the search strategy according to the strength of the coupling of the charged scalar to leptons. For $f\gtrsim  0.1$ we focus on the $3\ell$ channel, while for $f<0.1$ we make use of the $2\ell$ channel. In this manner we not only retain the sensitivity regardless of the order of magnitude of the coupling $f$ but also are able to translate readily the results obtained for the cross section into results for $\mathrm{BR}_{e+\mu}$ and/or $f$.  
%%%%%
%%%
%%%
%%%
%%%
%%%

Finally, we point out that the signal corresponding to the 2$\ell$ channel can be assumed to emerge from either the LL or LR operators, since they give rise to the same signature. This is not the case for the 3$\ell$ channel, as we will explain in Section~\ref{sec:3lep}. 
\section{Analysis in the two lepton channel}
\label{sec:2lep}
% Ref.~\cite{Alcaide:2017dcx}
In this Section we focus on the search strategy in the 2$\ell$ channel. The topology of the final state consists of two opposite-sign leptons and missing transverse energy. The relevant backgrounds are then Drell-Yan, $t\bar{t}$, $WW$, $WZ$, $ZZ$ and $tW$. We generated all these background processes at leading order for a center of mass energy of 13 TeV with {\tt MadGraph\_aMC@NLO~2.6}~\cite{Alwall:2014hca} and rescaled their cross sections with different $K$-factors to include the impact of QCD corrections. The parton shower and hadronization were carried out with {\tt PYTHIA 8}~\cite{Sjostrand:2014zea}, while the detector response was simulated with {\tt Delphes 3}~\cite{deFavereau:2013fsa}. In all the cases we impose the following set of cuts at generator level: $p^{\ell_1}_T>25$ GeV, $p^{\ell_2}_T>20$ GeV, $|\eta_{\ell}|<2.5$, where $\ell_{1}$ ($\ell_{2}$) denotes the leading (sub-leading) lepton. Additionally, for the Drell-Yan process we set $m_{\ell^+\ell^-}>100$~GeV in order to make the simulation more efficient. The information about the simulation of the different backgrounds is collected in Table~\ref{tab_bkg}. Regarding the event generation for the signal, we use the package \texttt{FeynRules}~\cite{Alloul:2013bka} to implement the relevant interactions and write them in the UFO format~\cite{Degrande:2011ua}. The rest of the simulation process proceeds in the same fashion as for the backgrounds. In particular, we demand the signal events to satisfy the same selection cuts at generator level as those imposed on the backgrounds.\par
%%%%%
\renewcommand{\arraystretch}{1.2}
\begin{table}[H]
\vspace*{4mm}
\begin{center}
\begin{tabular}{c|c|c|c}
\hline \hline
{\it Background} & {\it Cross section (pb)} & {\it K-factor} & {\it Simulated events} \\ \hline\hline 
$\mbox{Drell-Yan}$ & 81 & 1.2 & $5.0\times 10^{7}$ \\
$t\bar{t}$ & 20 & 1.8 &$2.5\times 10^{7}$ \\ 
$WW$ & 4.9 & 1.5&$3.0\times 10^{6}$ \\ 
$WZ$ & 2.0  & 1.4&$1.0\times 10^{6}$ \\ 
$ZZ$ & 0.8 & 1.4&$5.0\times 10^{5}$ \\
$tW$ & 4.2 & 0.9 &$1.5\times 10^6$ \\ \hline \hline
\end{tabular}
\caption{Main backgrounds along with their corresponding cross sections, the applied $K$-factors~\citep{Grazzini:2016ctr,Caola:2015rqy,Aad:2015zqe,Grazzini:2016swo,Boughezal:2016wmq,Ahrens:2011px,Czakon:2013goa,Cao:2008af,Frixione:2008yi,Kidonakis:2015nna} and the number of simulated events used in the analysis of the 2$\ell$-channel.}
\label{tab_bkg}
\end{center}
\end{table}
%%%%
At the reconstruction level we select events with two opposite-sign leptons%, where {\it leptons} refers to electrons or muons \ja{leptons ya esta definido mas arriba}
, that satisfy the same set of cuts imposed at generator level as well as the requirement $E^{\mathrm{miss}}_{T}>35$ GeV. The last cut on the total missing transverse energy is useful to reduce the Drell-Yan background which, unlike the signal, does not have large $E^{\mathrm{miss}}_T$. Out of this sample, we build the following three observables:
\begin{itemize}
\item[1.-] The invariant mass of the opposite-sign lepton pair, $m_{\ell\ell}$.
\item[2.-] The transverse mass of the opposite-sign lepton pair, $m_{T\ell\ell}$, defined by $$m^2_{T\ell\ell}=m^2_{\ell\ell}+2(E^{\ell\ell}_TE^{\mathrm{miss}}_T\,-\,\vec{p}^{\,\ell\ell}_T\cdot \vec{p}^{\,\,\mathrm{miss}}_T),$$ where $\vec{p}^{\,\ell\ell}_T=\vec{p}^{\,\,\ell^+}_T+\vec{p}^{\,\,\ell^-}_T$ and $E^{\ell\ell}_T=\sqrt{|\vec{p}^{\,\ell\ell}_T|^2+m^2_{\ell\ell}}$.
\item[3.-] The stransverse mass, $m_{T2}$, defined as 
\begin{equation}
\label{eqmT2}
m_{T2}=\mathrm{min}_{\vec{q}^{\,\mathrm{miss}}_{L1}+\vec{q}^{\,\mathrm{miss}}_{L2}=\vec{p}^{\,\mathrm{miss}}_T}\left\{\mathrm{max}\left[m_T(\vec{p}^{\,L1}_T,\vec{q}^{\,\,\mathrm{miss}}_{L1}\,),m_T(\vec{p}^{\,L2}_T,\vec{q}^{\,\,\mathrm{miss}}_{L2}\,)\right]\right\},
\end{equation}
with $m_T(\vec{p}^{\,X}_T,\vec{q}^{\,\,\mathrm{miss}}_{X}\,)=\sqrt{m^2_X+2(E^X_Tq^{\mathrm{miss}}_{X}-\vec{p}^{\,X}_T\cdot \vec{q}^{\,\,\mathrm{miss}}_{X}\,)}$, where $X$ denotes a visible particle, $E^X_T$ and $\vec{p}^{X}_T$ its transverse energy and momentum, respectively, while $\vec{q}^{\,\mathrm{miss}}_{X}$ is the part of the missing transverse momentum associated with $X$. %For the 2$\ell$ channel, t
The indices $L1$ and $L2$ stand for the harder and softer lepton, respectively, and we can safely set $m_X=0$ in the definition of $m_T$.
%%%
%\begin{equation}
%\label{eqmT2}
%m^2_{T2}=\mathrm{min}_{\not{\,\vec{q}_{1}}+\not{\,\vec{q}_{2}}=\vec{p}^{\,\mathrm{miss}}_T}\left\{\mathrm{max}\left[m_T(\vec{p}^{\,L1}_T,\not{\!\vec{q}_{1}}\,),m_T(\vec{p}^{\,L2}_T,\not{\!\vec{q}_{2}}\,)\right]\right\},
%\end{equation}
%with $m_T(\vec{p}^{\,X}_T,\not{\!\!\vec{q}}\,)=\sqrt{m^2_X+2(E^X_T|\!\!\not{\!\vec{q}}\,|-\vec{p}^{\,X}_T\cdot \not{\!\vec{q}}\,)}$, where $X$ denotes a visible particle and $E^X_T$ and $\vec{p}^{X}_T$ its tranverse energy and momentum, respectively. For the 2$\ell$ channel, the indices $L1$ and $L2$ stand for the harder and softer lepton, respectively, and we can safely set $m_X=0$ in the definition of $m_T$.
\end{itemize}
Additionally, we consider the observable $S_T$ defined as the scalar sum of the transverse momentum of all the leptons in the event. For each observable, $O=m_{\ell\ell},~m_{T\ell\ell}$ and $m_{T2}$, we build 81 different categories determined by the requirements $O>X$ and $S_T>Y$, where $X,Y=100, 200,\cdots, 900$~GeV. Since the $m_{T2}$ variable is obtained from the transverse masses corresponding to the two leptons arising from the decay of $h^+$ and $h^-$, its distribution exhibits an endpoint around the charged scalar mass. Therefore, the lowest cut value chosen for this observable in the definition of the categories (100 GeV) is not appropriate for values of $m_{h^{\pm}}$ close to it. For this reason, we added 18 categories corresponding to $m_{T2}>70$~GeV and $m_{T2}>80$~GeV (with $S_T > X$ and $X=100,...,900$ GeV).\par
\begin{figure}[H]
\begin{center}
\includegraphics[scale=0.55]{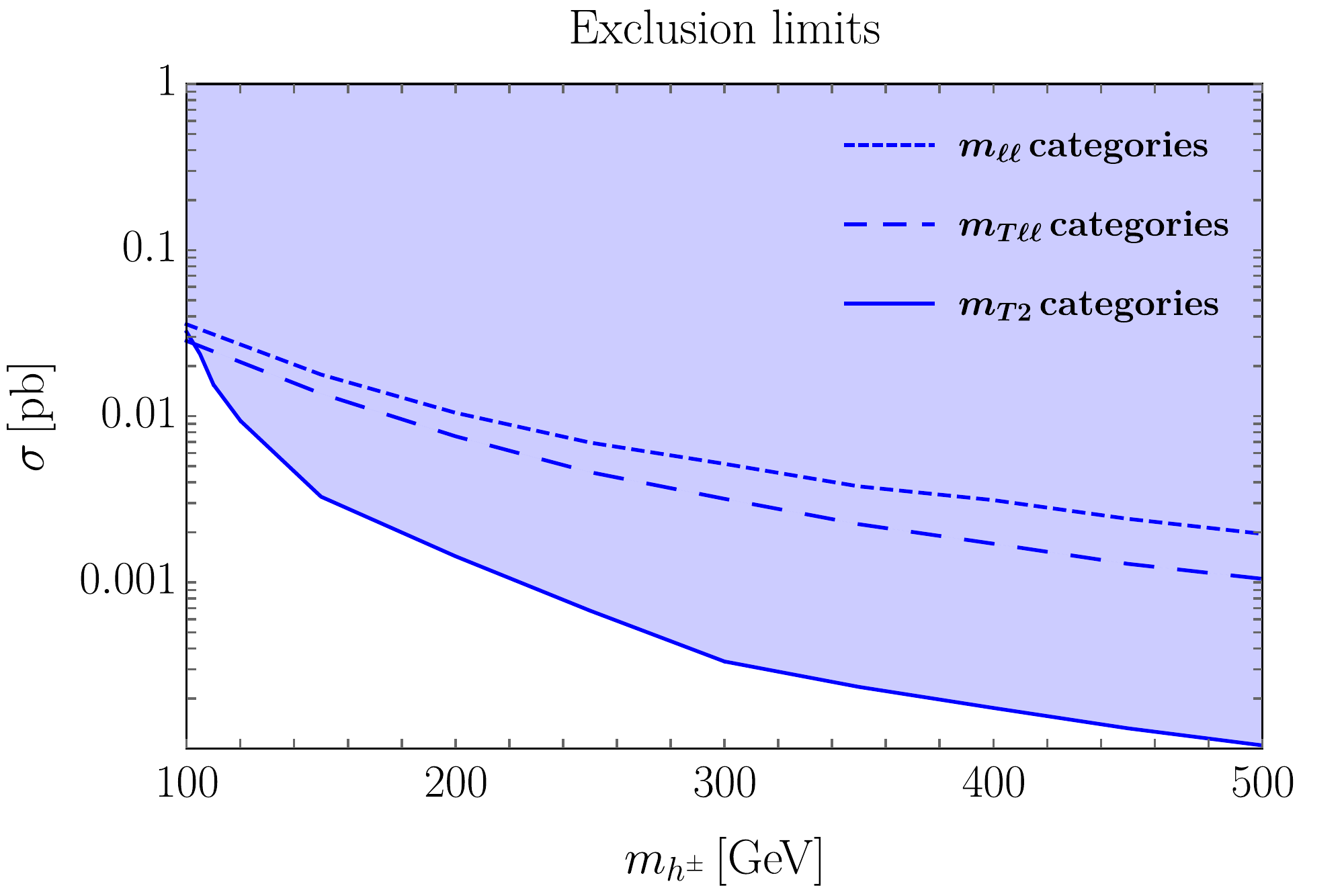} 
\caption{Bounds on the cross section of pair produced $h^{\pm}$ decaying into electrons and muons for a total integrated luminosity of 300 $\mathrm{fb}^{-1}$.}
\label{figxslimit2l}
\end{center}
\end{figure}
\begin{figure}[H]
\begin{center}
\includegraphics[scale=0.55]{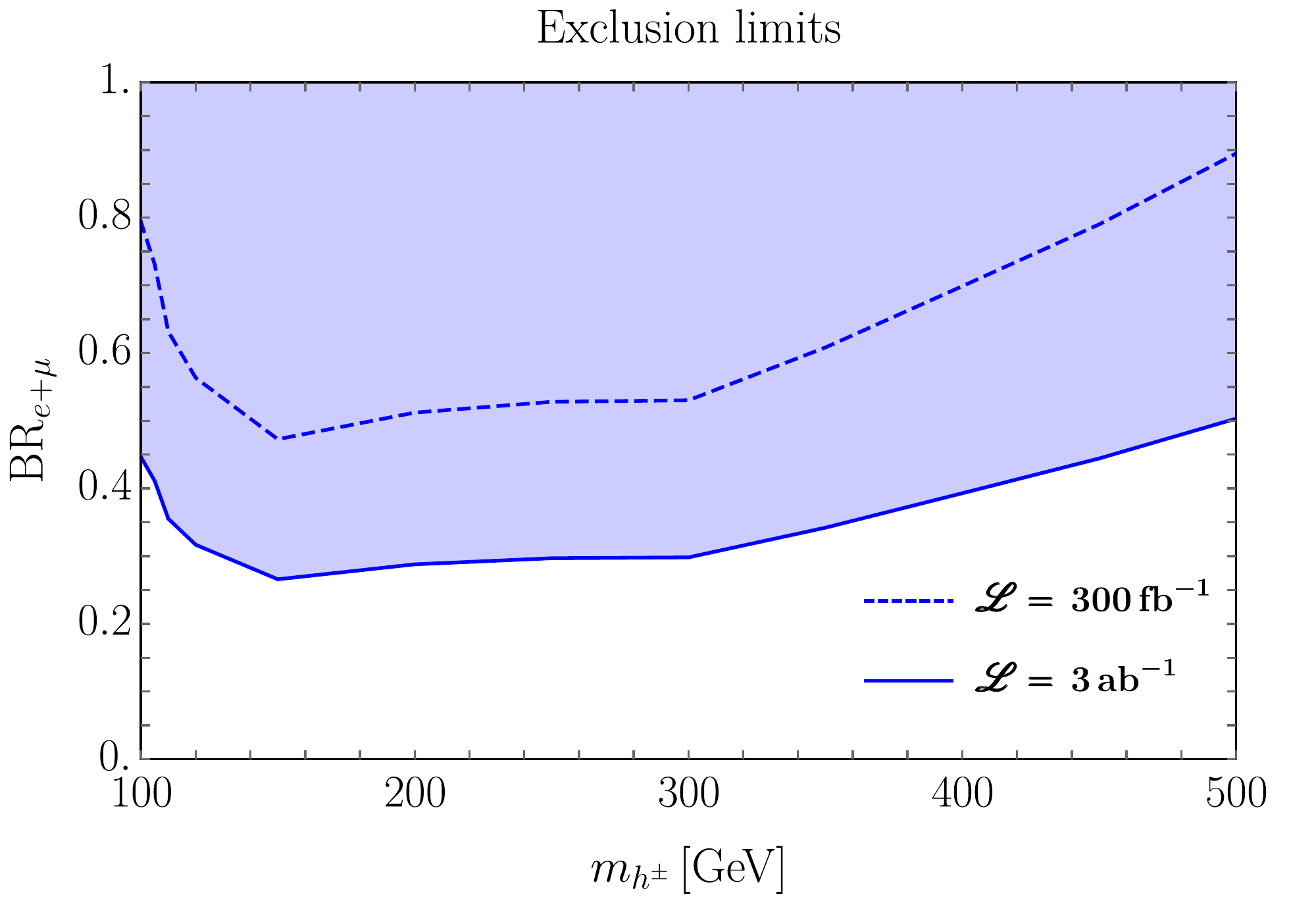} 
\caption{Bounds in the decay branching ratio of $h^{\pm}$ into electrons and muons obtained by using the $m_{T2}$ categories with luminosities of 300 $\mathrm{fb}^{-1}$ (dashed line) and 3 $\mathrm{ab}^{-1}$ (solid line).}
\label{figBRlimit2l}
\end{center}
\end{figure}

%%%%
We vary the charged scalar mass, $m_{h^{\pm}}$, between 100 and 500 GeV in steps of 50 GeV and for each value we estimate the lower cross section that can be excluded with a luminosity of 300~$\mathrm{fb}^{-1}$ by using the $m_{\ell\ell},~m_{T\ell\ell}$ and $m_{T2}$ categories. Exclusions for intermediate masses are obtained by linear interpolation. The results are shown in Figure~\ref{figxslimit2l}. The sensitivity is driven by the $m_{T2}$ categories in all the considered range of masses, even when it worsens significantly below $m_{h^{\pm}}\sim 150$ GeV and becomes similar to that achieved with the other observables around $m_{h^{\pm}}=100$ GeV. With the $m_{T2}$ categories it is possible to exclude cross sections ranging from $\sim$30 fb to 0.1 fb for masses between 100 GeV and 500 GeV. 
It is interesting to interpret the limits given in Figure~\ref{figxslimit2l} in terms of the decay branching ratio of $h^{\pm}$ into electrons and muons. This is easy to accomplish by remembering from Sec.~\ref{sec:motivation} that the signal cross section factorizes as $\sigma(pp\to h^+h^-)\times\mathrm{BR}_{e+\mu}^2$. We display the results in Figure~\ref{figBRlimit2l} for the most sensitive observable ($m_{T2}$) and two values of the total integrated luminosity, $\mathscr{L}=300\, \mathrm{fb}^{-1}$ and $3\, \mathrm{ab}^{-1}$.  We see from the figure that charged scalars decaying mostly to electrons and muons ($\mathrm{BR}_{e+ \mu}\sim 0.9$) can be excluded up to 500 GeV with 300 $\mathrm{fb}^{-1}$. By increasing the luminosity to 3 $\mathrm{ab}^{-1}$ this conclusion can be extended for charged scalars with $\mathrm{BR}_{e+\mu} \geqslant 0.5$. For a charged scalar decaying only into leptons, the sensitivity gap for $\mathrm{BR}_{e+\mu}<0.3$ could be addressed in principle by considering the ditau channel since $\mathrm{BR}_{\tau}=1-\mathrm{BR}_{e+\mu}$ and then the exclusion limits in terms of $\mathrm{BR}_{\tau}$ translate into an upper limit on $\mathrm{BR}_{e+ \mu}$. In fact, by combining our results with the exclusion reported in~\cite{Cao:2017ffm} which is based on the recasting of the analysis of Ref.~\cite{CMS:2017rio}, we can exclude singlet charged scalars with masses below $\sim 280$ GeV. A search strategy in the ditau channel more focused on the high-mass range could extend this exclusion, however this is beyond the approach of this paper and is left as future work. Finally, a charged scalar decaying fully into electrons and muons could be ruled out in all the considered mass range (100~GeV - 500~GeV) with a minimum luminosity of $\sim 192\,\mathrm{fb}^{-1}$.
%%%%%
%%%%%
\section{Analysis in the three lepton channel}
\label{sec:3lep}
As previously discussed in Sec.~\ref{sec:motivation}, the $3\ell$ channel is more suitable to probe the large Yukawa coupling regime. In order to develop the search strategy in this channel we make use of the background presented in Ref.~\cite{Alcaide:2017dcx}, where a number of events consistent with an integrated luminosity of $3$ ab$^{-1}$ was generated at $\sqrt s = 13$ TeV. The relevant backgrounds consist of $WZ$, $ZZ$, $WWW$, $WWZ$, $WZZ$, $ZZZ$, $ttW$ and $ttZ$. 
At the reconstruction level we demand electrons (muons) to have $p_{T} > 20$~$(10)$ GeV and $|\eta| < 2.5$ $(2.6)$. 
We construct the following two observables\footnote{We do not consider the observable $m_{T2}$ since, unlike in the $2\ell$ channel, there is only one source of missing energy in this case. Moreover, in Ref.~\cite{Alcaide:2017dcx} this observable was constructed taking $L1$ as the vectorial sum of the two same-sign leptons, while $L2$ was given by the third one. For this choice, $m_X$ in the definition of $m_{T2}$ (see Eq.~(\ref{eqmT2})) cannot be neglected, as it corresponds to the invariant mass of the two same-sign leptons. This will be enough to make most $m_{\ell\ell}$ and $m_{T2}$ categories almost identical.}:
\begin{itemize}
\item[1.-]  $m_{\ell\ell}$, the invariant mass of the two same-sign leptons, and
\item[2.-]   the transverse mass, defined in the previous section, of the same-sign lepton pair ($m_{T\ell\ell}$), as well as the one of the third lepton ($m_{T\ell}$).
\end{itemize}
Like in Sec.~\ref{sec:2lep}, we additionally consider the auxiliary observable $S_T$. Once again, for each observable $O = m_{\ell\ell}$, $m_T$ we build 81 different categories with $O >X$ and $S_T > Y$, where $X, Y = 100,200,...,900$~GeV. We remark that, contrary to the pair production channel, in this case there are two transverse masses and we demand both of them to simultaneously fulfil the selection cuts. From now on we will often speak of $m_T$ analysis to refer to the combined analysis of the two transverse masses.

%We do not consider the observable $m_{T2}$ since, unlike in the $2l$ channel, there is only one source of missing energy. Moreover, in paper of Ref.~\cite{Alcaide:2017dcx} this observable was constructed taking $L1$ as the vectorial sum of the two same-sign leptons, while $L2$ was given by the third one. For this choice, $m_X^2$ in the above definition of $m_{T2}$ cannot be neglected, as it is identified with the invariant mass of the two same-sign leptons. This will be enough to make most $m_{ll}$ and $m_{T2}$ categories almost identical. 
%

Regarding the signal events, we use the package \texttt{FeynRules}~\cite{Alloul:2013bka} to generate the UFO~model~\cite{Degrande:2011ua} that implements the relevant interactions. In order to be consistent with the generation of background events in Ref.~\cite{Alcaide:2017dcx}, we choose once again \texttt{MadGraph 5}~\cite{Alwall:2014hca} and then \texttt{PYTHIA 6}~\cite{ref_pythia6} for parton showering. We demand the signal events to satisfy the same selection cuts as the background. 

We consider the mass of the charged scalar singlet to range between 100 and 500 GeV, in steps of 50 GeV. 
%\ja{maybe say that the production depends (almost) exclusively on $f_{12}$ so we only need to set this coupling. But this should have been explained in more detail in section II. The other two are set to zero so that we have BR=1}. 
%Hereafter we set $f=1$ in all simulations, as the results can be rescaled to any coupling strength following eq. 2. \ja{link equation}
Considering solely electrons and muons (plus missing transverse energy) in the final state, the production is such that only the coupling $f_{e\mu}$ is needed.
%\nm{queda claro que todo el proceso depende solo de $f_{e\mu}$ y no solo la produccion?}. 
Hereafter we set $f_{e\mu}=1$ in all the simulations, as the results can be rescaled to any coupling strength following Eq.~(\ref{eqscale3lep}). %The remaining two couplings are set to zero so that BR$=1$.
For each observable and every value of the scalar mass we look for the category with the largest sensitivity. By following this procedure, we compute the lowest cross section that can be excluded at the 95\% C.L. Exclusions for intermediate masses are obtained by linear interpolation. For $m_{h^\pm}\gtrsim250$ GeV the sensitivity is slightly driven by $m_T$, while for lower values of the mass, the sensitivity of the transverse mass worsens due to the presence of an endpoint in its distribution around the scalar mass. This makes the observable $m_{ll}$ the one with the best sensitivity in that region.

\begin{figure}[H]
\centering
\includegraphics[width=0.9\textwidth]{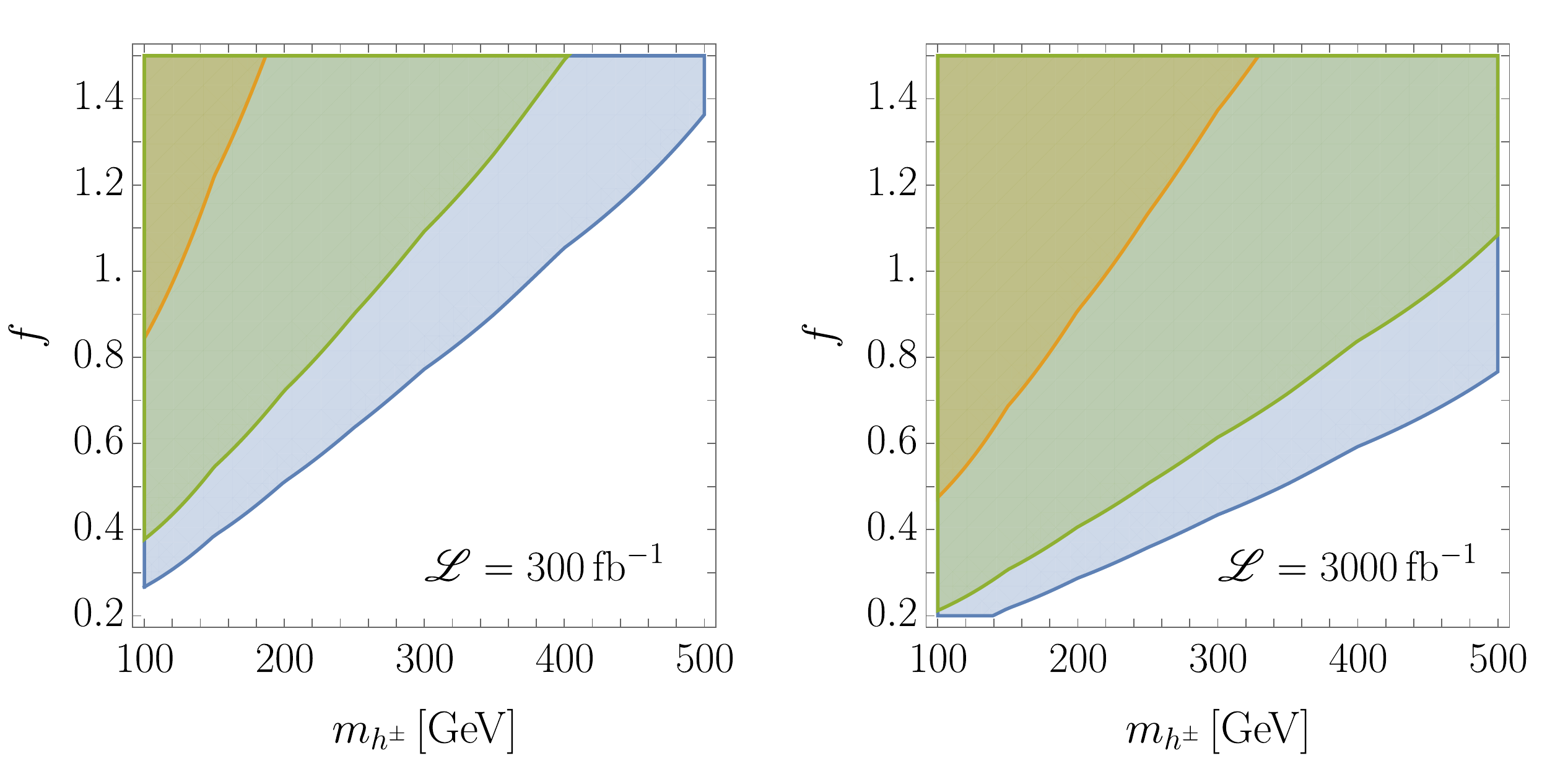}
\caption{Exclusion limits at 95\% C.L. of the Yukawa coupling of $h^{\pm}$ decaying into electrons and muons in the $3l$ channel for a total integrated luminosity of 300 fb$^{-1}$ (left panel) and 3 ab$^{-1}$ (right panel). Colors correspond to: $\text{BR}_{e+\mu}=0.1$ (orange), $\text{BR}_{e+\mu}=0.5$ (green) and $\text{BR}_{e+\mu}=1$ (blue).}
\label{fig:gm_3l}
\end{figure}

\begin{figure}[H]
\centering
\subfloat[LL interaction]{\includegraphics[align=c,scale=0.45]{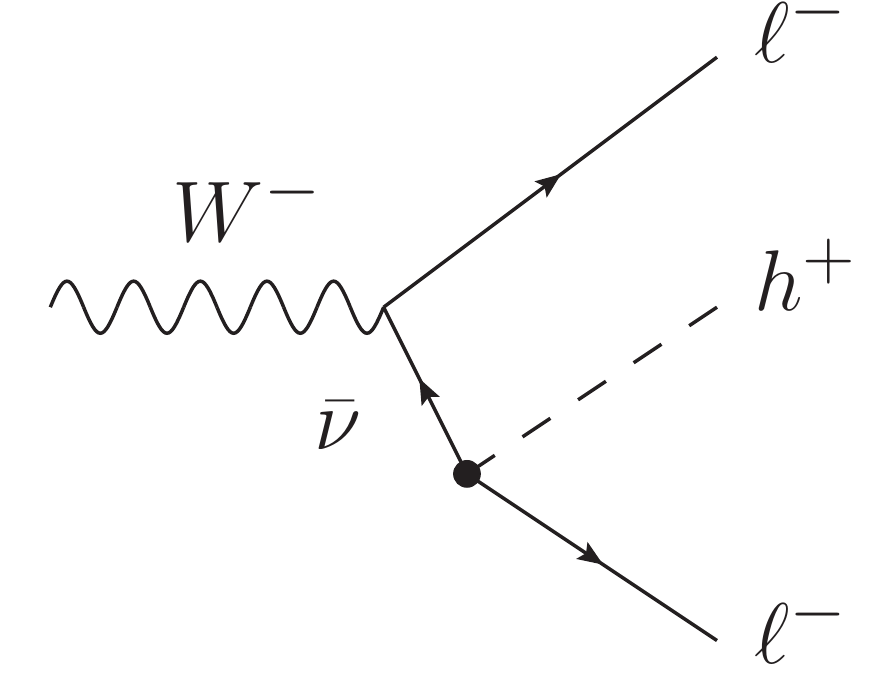}}\hspace*{25mm}
\subfloat[LR interaction]{\includegraphics[align=c,scale=0.45]{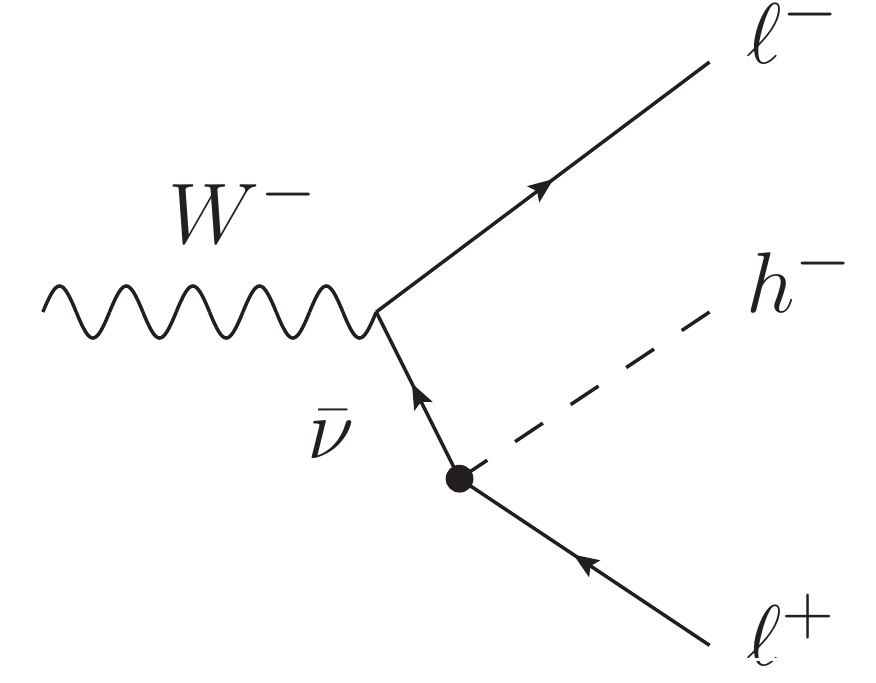}}
\caption{Feynman diagrams representing the production of the charged scalar through the LL operator~(panel~a) and the LR operator (panel~b).}
\label{fig:feymandiag2}
\end{figure}

\begin{figure}[H]
\centering
\includegraphics[width=1.0\textwidth]{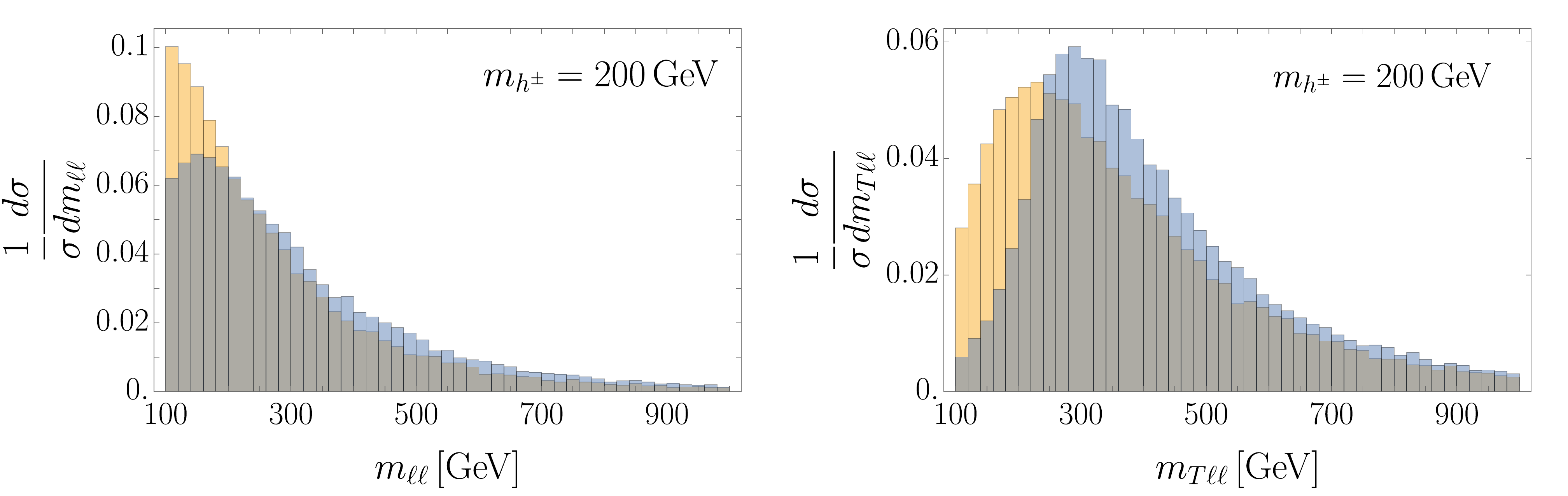}
\includegraphics[width=0.5\textwidth]{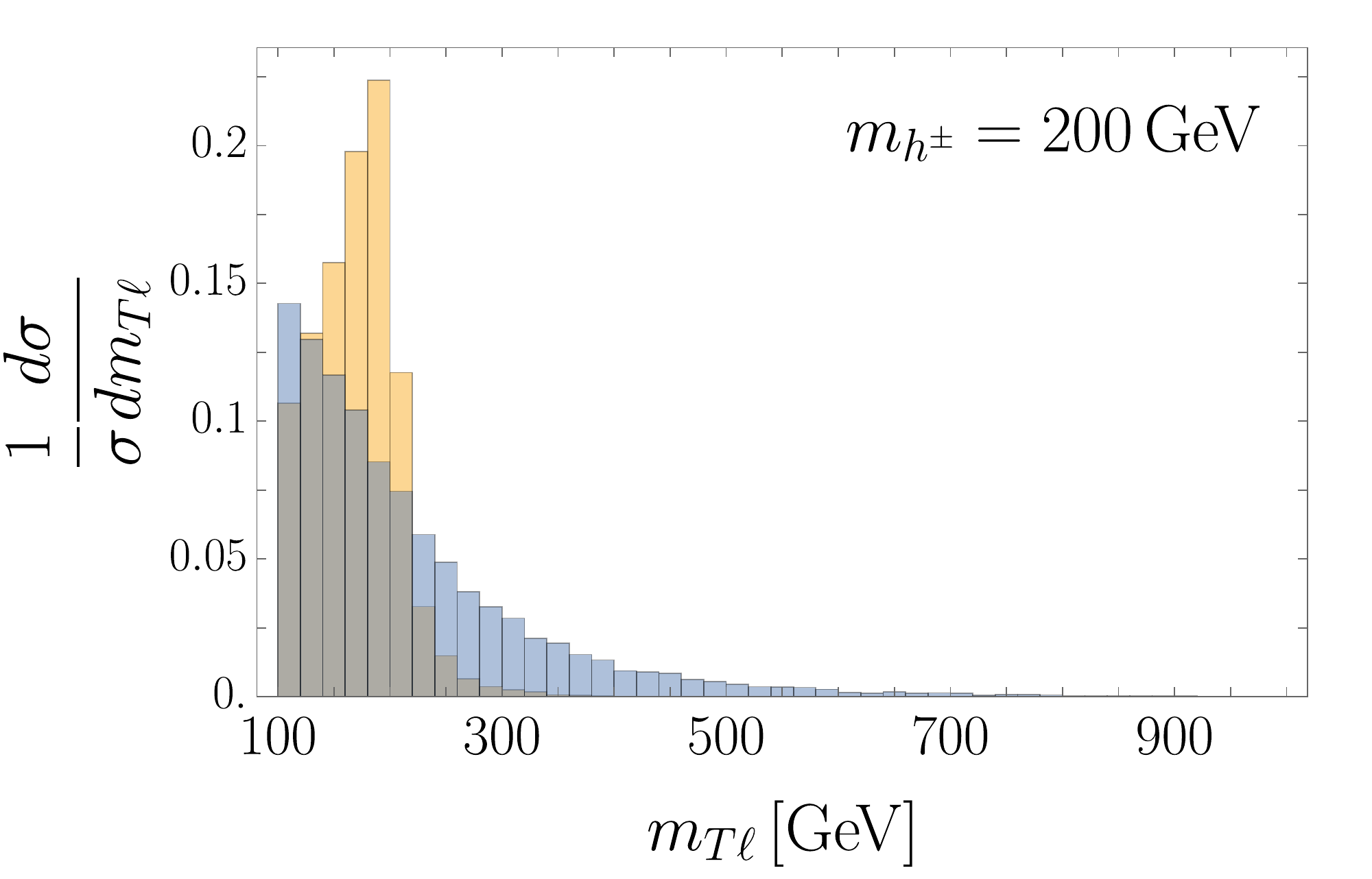}
\caption{Top left panel: Invariant mass distribution of two same-sign leptons in events generated with the LL~coupling (orange) and the LR interaction (blue), for  $m_{h^\pm}=200$ GeV. Top right panel: same as before but for the transverse mass distribution of two same-sign leptons. Bottom panel: same as before but for the transverse mass distribution of the single opposite-sign lepton. }
\label{fig:histograms}
\end{figure}

The results are shown in Figure \ref{fig:gm_3l} for a total integrated luminosity of 300 fb$^{-1}$~(left panel) and 3~ab$^{-1}$~(right panel). Making use of Eq.~(\ref{eqscale3lep}), we plotted the coupling-mass plane for three different decay branching ratios of the charged scalar, namely $\text{BR}_{e+\mu}=0.1$~(orange), $\text{BR}_{e+\mu}=0.5$~(green) and $\text{BR}_{e+\mu}=1$~(blue). If $h^\pm$ decays only to electrons and muons, it is possible to exclude the entire range of studied masses for couplings larger than $\sim1.4$ with a luminosity of 300 fb$^{-1}$. For the high luminosity phase, with 3 ab$^{-1}$, this limit could be extended to couplings as low as $\sim 0.8$. Additionally, smaller branching ratios become more accessible. For instance, a charged scalar with BR$_{e+\mu}=0.5$ could be probed up to $m_{h}\sim500$~GeV with a coupling $\sim 1.1$. 
%Since the total cross section depends quadratically on the Yukawa coupling (see eq. \ref{eqscale3lep}), exclusion limits on lower couplings become harder to probe.
%\nm{de la seccion 2 dimos la conclusion de que por encima de $\approx$0.1 convenia utilizar este canal, de manera que quiza sea bueno enfatizar que con solo 300 fb$^{-1}$ podemos excluir la mayor parte del rango donde el canal es util, y que con 3000fb$^{-1}$ se excluye completamente (el 0.1 era aproximado, con el limite de 0.15 estamos cubriendo efectivamenete todo el rango para el cual la busqueda en este canal fue pensada)}.  
%%%
%%%
Contrary to the $2\ell$-channel, in the $3\ell$-channel the analysis based on the LR interaction will report, to some extent, different results than the one considering the LL operator.
In particular, the topology with the LR interaction shows a different configuration of the electric charges of the three leptons in the final state (see Figure \ref{fig:feymandiag2}). 
For this reason the observables are constructed with different leptons, and we expect the distributions to disagree from one scenario to the other.
In Figure \ref{fig:histograms} we show different distributions in both frameworks, for a mass of 200~GeV. 
Notably and in contrast to the case with the LL interaction, $m_{T\ell}$ does not show an endpoint around the mass of the charged scalar since it is not built with the lepton arising from the decay of the scalar. This feature could help to distinguish one scenario from the other.
Additionally, while with the LL operator two diagrams contribute to the amplitude, there is only one diagram in the LR topology, making its production cross section typically one order of magnitud smaller. 

\begin{figure}[H]
\centering
\includegraphics[width=0.65\textwidth]{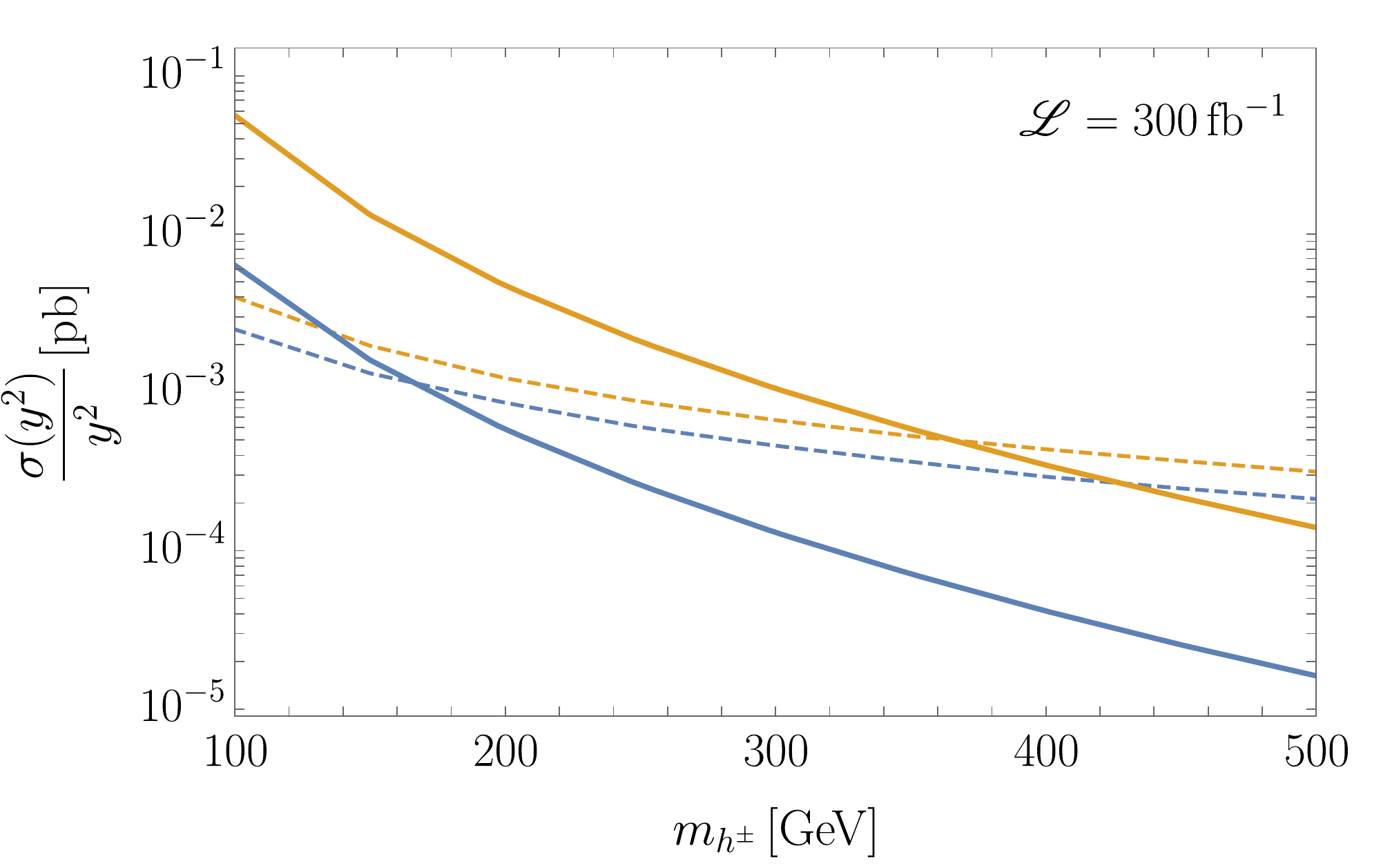}
\caption{Bounds on the cross section of a singly $h^\pm$ produced through the LR interaction (dashed blue line) and the LL operator (dashed orange line) for a luminosity of $L = 300$ fb$^{-1}$, and decaying exclusively to electrons and muons. The theoretical cross section in the LR (solid blue line) and in the LL (solid orange line) case are also shown for reference. The coupling $y$ represents the Yukawa couplings $g$ ($f$) in the LR (LL) scenario.}
\label{fig:xsRenormalizableVSeffective}
\end{figure}

In order to illustrate to what extent the change on the kinematic distributions impacts on the results, let us assume that the Lagrangian in Eq.~(\ref{eq:LRlagrangian}) %also
 preserves lepton flavour (that is, we only keep the diagonal elements) and redo the whole procedure, fixing the couplings $g_{ee} = g_{\mu\mu}$ for simplicity.� Results for the LR operator are depicted in blue in Figure \ref{fig:xsRenormalizableVSeffective} for a luminosity of 300~fb$^{-1}$, in comparison to those obtained with the LL interaction in orange. The symbol $y$ denotes the Yukawa coupling $g$ ($f$) in the LR (LL) framework. Solid lines describe the theoretical cross section while dashed lines indicate the maximum cross section that can be probed at the given luminosity.
The latter strongly depends on the acceptance of each category and observable. In the whole interval of study, the sensitivity for the LR operator is driven by $m_{ll}$, whose distribution is very similar to that of the LL interaction. Conversely, the theoretical cross section, as explained above, is typically one order of magnitude smaller. 
For this reason the exclusion limits will be weaker in this scenario and for retrieving the same sensitivities achieved with the LL framework, larger couplings or luminosities are needed. 
In particular, a $h^\pm$ decaying exclusively to electrons and muons through the LR interaction can be excluded with a mass as large as $\sim170$ GeV with a luminosity of 300 fb$^{-1}$, whereas this limit extends to $\sim 370$ GeV for the LL interaction.

\section{Conclusions}
\label{sec:conclusions}
%Other searches of particles with similar quantum numbers, as the ones of sleptons, focus the analysis in pair-produced charged scalars decaying into charged leptons and missing transverse energy. 
In this paper we have proposed a search strategy for singly-charged scalars focusing in final states with electrons and muons. We have shown that the single production mode through a virtual lepton exchange may have a significant impact on the kinematic distributions in a signature with two leptons for large Yukawa couplings, and then it should be added to the usual pair production mode. Taking this into account, we have split our analysis according to the strength of the interactions of the charged scalar with leptons. For large Yukawa couplings, we have considered a final state consisting of $3\ell + E^{\mathrm{miss}}_T$ which arises from the single production of the charged scalar and its subsequent decay into a lepton and a neutrino. For small Yukawa couplings, the $3\ell$ channel loses its sensitivity and then we have considered instead the topology $2\ell + E^{\mathrm{miss}}_T$ since the corresponding cross section is dominated in this scenario by the pair production channel which is completely determined by the gauge couplings of the charged scalar. In this manner, each topology is driven by only one production mode. 
%We have pointed out that in the case of a signature with two leptons and missing transverse energy, the single production mechanism through a virtual lepton exchange may have a significant impact on the kinematic distributions and then it should be taken into account in addition to the usual pair production mode
%Interactions of charged scalar with a charged lepton and a neutrino are described with operators of both types LL and LR. The simplest way to realised the operator LL is with the charged scalar being a weak singlet. 
%It renormalizable interaction with leptons seemingly induces large lepton flavour violation which are deeply constrained by lepton rare decay data. Nevertheless there are ways to escape those bounds, such as imposing some type of global lepton number symmetry. Other possibilities for generating this operator as well as the LR are to consider the charged scalar as a member of a larger multiplet or as a singlet in a EFT scheme. We have explored the collider phenomenology of a charged scalar that couples to leptons applicable to every one of these frameworks. 

Within the low Yukawa coupling scenario, we have found that a charged scalar decaying exclusively into electrons and muons can be excluded up to masses of $500$ GeV with an integrated luminosity of $\sim 200$ fb$^{-1}$. In the high luminosity phase it would be possible to exclude all the studied mass range for any charged scalar with $\mathrm{BR}_{e+\mu}\geqslant 0.5$. In the large coupling scenario (coupling $\sim1$), the same range of masses can be accessed with a luminosity of $\sim 300$ fb$^{-1}$ for couplings $\gtrsim 1.4$, while in the high luminosity phase this limit could be extended to couplings as low as $\sim 0.8$.
Although our results were obtained by assuming that the singly-charged scalar is a weak singlet with renormalizable interactions, in principle they can also be applied to other models of charged scalars with similar production and decay modes. However, we have shown that special attention is needed when using the results obtained in the $3\ell$ channel since in this case the search strategy is sensitive to the type of interaction of the charged scalar (LL or LR). By recasting our search strategy for the LR interaction we concluded that the sensitivity decreases in this scenario. In particular, for couplings of order one, a charged scalar decaying fully into electrons and muons via the LL interaction can be excluded up to a mass of $\sim 370$ GeV, while this limit relaxes to $\sim 170$ GeV if the decay proceeds through the LR interaction.
\section*{Acknowledgments}
We thank Mikael Chala and Arcadi Santamaria for useful discussions in early stages of this paper. N.M. would like to thank Alejandro Szynkman for fruitful discussions regarding the experimental constraints.
% and a careful reading.
All Feynman diagrams have been drawn with JaxoDraw \cite{ref:jaxodraw1,ref:jaxodraw2}. This work has been partially supported by CONICET and ANPCyT under projects PICT 2016-0164 and PICT 2017-2751 (N. M.) and by the Spanish MINECO under grant FPA2014-54459-P (J. A.).
%%%%%%%%%%
\section*{\refname}
\let\bibsection\relax
\makeatletter
\newcommand{\bibnote}[2]{\@namedef{#1note}{#2}}
\makeatother
\bibliography{biblio}

\begin{thebibliography}{47}%
\makeatletter
\providecommand \@ifxundefined [1]{%
 \@ifx{#1\undefined}
}%
\providecommand \@ifnum [1]{%
 \ifnum #1\expandafter \@firstoftwo
 \else \expandafter \@secondoftwo
 \fi
}%
\providecommand \@ifx [1]{%
 \ifx #1\expandafter \@firstoftwo
 \else \expandafter \@secondoftwo
 \fi
}%
\providecommand \natexlab [1]{#1}%
\providecommand \enquote  [1]{``#1''}%
\providecommand \bibnamefont  [1]{#1}%
\providecommand \bibfnamefont [1]{#1}%
\providecommand \citenamefont [1]{#1}%
\providecommand \href@noop [0]{\@secondoftwo}%
\providecommand \href [0]{\begingroup \@sanitize@url \@href}%
\providecommand \@href[1]{\@@startlink{#1}\@@href}%
\providecommand \@@href[1]{\endgroup#1\@@endlink}%
\providecommand \@sanitize@url [0]{\catcode `\\12\catcode `\$12\catcode
  `\&12\catcode `\#12\catcode `\^12\catcode `\_12\catcode `\%12\relax}%
\providecommand \@@startlink[1]{}%
\providecommand \@@endlink[0]{}%
\providecommand \url  [0]{\begingroup\@sanitize@url \@url }%
\providecommand \@url [1]{\endgroup\@href {#1}{\urlprefix }}%
\providecommand \urlprefix  [0]{URL }%
\providecommand \Eprint [0]{\href }%
\providecommand \doibase [0]{https://doi.org/}%
\providecommand \selectlanguage [0]{\@gobble}%
\providecommand \bibinfo  [0]{\@secondoftwo}%
\providecommand \bibfield  [0]{\@secondoftwo}%
\providecommand \translation [1]{[#1]}%
\providecommand \BibitemOpen [0]{}%
\providecommand \bibitemStop [0]{}%
\providecommand \bibitemNoStop [0]{.\EOS\space}%
\providecommand \EOS [0]{\spacefactor3000\relax}%
\providecommand \BibitemShut  [1]{\csname bibitem#1\endcsname}%
\let\auto@bib@innerbib\@empty
%</preamble>
\bibitem [{\citenamefont {Branco}\ \emph {et~al.}(2012)\citenamefont {Branco},
  \citenamefont {Ferreira}, \citenamefont {Lavoura}, \citenamefont {Rebelo},
  \citenamefont {Sher},\ and\ \citenamefont {Silva}}]{Branco}%
  \BibitemOpen
  \bibfield  {author} {\bibinfo {author} {\bibfnamefont {G.~C.}\ \bibnamefont
  {Branco}}, \bibinfo {author} {\bibfnamefont {P.~M.}\ \bibnamefont
  {Ferreira}}, \bibinfo {author} {\bibfnamefont {L.}~\bibnamefont {Lavoura}},
  \bibinfo {author} {\bibfnamefont {M.~N.}\ \bibnamefont {Rebelo}}, \bibinfo
  {author} {\bibfnamefont {M.}~\bibnamefont {Sher}},\ and\ \bibinfo {author}
  {\bibfnamefont {J.~P.}\ \bibnamefont {Silva}},\ }\bibfield  {title} {\bibinfo
  {title} {{\it Theory and phenomenology of two-Higgs-doublet models}},\ }\href
  {https://doi.org/10.1016/j.physrep.2012.02.002} {\bibfield  {journal}
  {\bibinfo  {journal} {Phys. Rept.}\ }\textbf {\bibinfo {volume} {516}},\
  \bibinfo {pages} {1} (\bibinfo {year} {2012})},\ \Eprint
  {https://arxiv.org/abs/1106.0034} {arXiv:1106.0034 [hep-ph]} \BibitemShut
  {NoStop}%
%%CITATION = ARXIV:1106.0034;%%
\bibitem [{\citenamefont {Abbiendi}\ \emph {et~al.}(2013)\citenamefont
  {Abbiendi} \emph {et~al.}}]{LEP}%
  \BibitemOpen
  \bibfield  {author} {\bibinfo {author} {\bibfnamefont {G.}~\bibnamefont
  {Abbiendi}} \emph {et~al.} (\bibinfo {collaboration} {ALEPH, DELPHI, L3,
  OPAL, LEP}),\ }\bibfield  {title} {\bibinfo {title} {{\it Search for Charged
  Higgs bosons: Combined Results Using LEP Data}},\ }\href
  {https://doi.org/10.1140/epjc/s10052-013-2463-1} {\bibfield  {journal}
  {\bibinfo  {journal} {Eur. Phys. J.}\ }\textbf {\bibinfo {volume} {C73}},\
  \bibinfo {pages} {2463} (\bibinfo {year} {2013})},\ \Eprint
  {https://arxiv.org/abs/1301.6065} {arXiv:1301.6065 [hep-ex]} \BibitemShut
  {NoStop}%
%%CITATION = ARXIV:1301.6065;%%
\bibitem [{\citenamefont {Aad}\ \emph {et~al.}(2013)\citenamefont {Aad} \emph
  {et~al.}}]{cs}%
  \BibitemOpen
  \bibfield  {author} {\bibinfo {author} {\bibfnamefont {G.}~\bibnamefont
  {Aad}} \emph {et~al.} (\bibinfo {collaboration} {ATLAS}),\ }\bibfield
  {title} {\bibinfo {title} {{\it Search for a light charged Higgs boson in the
  decay channel $H^+ \to c\bar{s}$ in $t\bar{t}$ events using pp collisions at
  $\sqrt{s}$ = 7 TeV with the ATLAS detector}},\ }\href
  {https://doi.org/10.1140/epjc/s10052-013-2465-z} {\bibfield  {journal}
  {\bibinfo  {journal} {Eur. Phys. J.}\ }\textbf {\bibinfo {volume} {C73}},\
  \bibinfo {pages} {2465} (\bibinfo {year} {2013})},\ \Eprint
  {https://arxiv.org/abs/1302.3694} {arXiv:1302.3694 [hep-ex]} \BibitemShut
  {NoStop}%
%%CITATION = ARXIV:1302.3694;%%
\bibitem [{\citenamefont {Aaboud}\ \emph
  {et~al.}(2018{\natexlab{a}})\citenamefont {Aaboud} \emph {et~al.}}]{taunu}%
  \BibitemOpen
  \bibfield  {author} {\bibinfo {author} {\bibfnamefont {M.}~\bibnamefont
  {Aaboud}} \emph {et~al.} (\bibinfo {collaboration} {ATLAS}),\ }\bibfield
  {title} {\bibinfo {title} {{\it Search for charged Higgs bosons decaying via
  $H^{\pm} \to \tau^{\pm}\nu_{\tau}$ in the $\tau$+jets and $\tau$+lepton final
  states with 36 fb$^{-1}$ of $pp$ collision data recorded at $\sqrt{s} = 13$
  TeV with the ATLAS experiment}},\ }\href
  {https://doi.org/10.1007/JHEP09(2018)139} {\bibfield  {journal} {\bibinfo
  {journal} {JHEP}\ }\textbf {\bibinfo {volume} {09}},\ \bibinfo {pages}
  {139}},\ \Eprint {https://arxiv.org/abs/1807.07915} {arXiv:1807.07915
  [hep-ex]} \BibitemShut {NoStop}%
%%CITATION = ARXIV:1807.07915;%%
\bibitem [{\citenamefont {Aaboud}\ \emph
  {et~al.}(2018{\natexlab{b}})\citenamefont {Aaboud} \emph {et~al.}}]{tb1}%
  \BibitemOpen
  \bibfield  {author} {\bibinfo {author} {\bibfnamefont {M.}~\bibnamefont
  {Aaboud}} \emph {et~al.} (\bibinfo {collaboration} {ATLAS}),\ }\bibfield
  {title} {\bibinfo {title} {{\it Search for charged Higgs bosons decaying into
  top and bottom quarks at $\sqrt{s}$ = 13 TeV with the ATLAS detector}},\
  }\href {https://doi.org/10.1007/JHEP11(2018)085} {\bibfield  {journal}
  {\bibinfo  {journal} {JHEP}\ }\textbf {\bibinfo {volume} {11}},\ \bibinfo
  {pages} {085}},\ \Eprint {https://arxiv.org/abs/1808.03599} {arXiv:1808.03599
  [hep-ex]} \BibitemShut {NoStop}%
%%CITATION = ARXIV:1808.03599;%%
\bibitem [{\citenamefont {{ATLAS collaboration}}(2016)}]{tb2}%
  \BibitemOpen
  \bibfield  {author} {\bibinfo {author} {\bibnamefont {{ATLAS
  collaboration}}},\ }\href {https://cds.cern.ch/record/2206809} {\emph
  {\bibinfo {title} {{\it Search for charged Higgs bosons in the $H^{\pm}\to
  tb$ decay channel in $pp$ collisions at $\sqrt{s}=13$ TeV using the ATLAS
  detector}}}},\ \bibinfo {type} {Tech. Rep.}\ \bibinfo {number}
  {ATLAS-CONF-2016-089}\ (\bibinfo  {institution} {CERN},\ \bibinfo {address}
  {Geneva},\ \bibinfo {year} {2016})\BibitemShut {NoStop}%
\bibitem [{\citenamefont {Georgi}\ and\ \citenamefont
  {Machacek}(1985)}]{Georgi}%
  \BibitemOpen
  \bibfield  {author} {\bibinfo {author} {\bibfnamefont {H.}~\bibnamefont
  {Georgi}}\ and\ \bibinfo {author} {\bibfnamefont {M.}~\bibnamefont
  {Machacek}},\ }\bibfield  {title} {\bibinfo {title} {{\it Doubly charged
  Higgs bosons}},\ }\href
  {https://doi.org/https://doi.org/10.1016/0550-3213(85)90325-6} {\bibfield
  {journal} {\bibinfo  {journal} {Nucl. Phys.}\ }\textbf {\bibinfo {volume}
  {B262}},\ \bibinfo {pages} {463 } (\bibinfo {year} {1985})}\BibitemShut
  {NoStop}%
\bibitem [{\citenamefont {Magg}\ and\ \citenamefont
  {Wetterich}(1980)}]{seesawII1}%
  \BibitemOpen
  \bibfield  {author} {\bibinfo {author} {\bibfnamefont {M.}~\bibnamefont
  {Magg}}\ and\ \bibinfo {author} {\bibfnamefont {C.}~\bibnamefont
  {Wetterich}},\ }\bibfield  {title} {\bibinfo {title} {{\it Neutrino mass
  problem and gauge hierarchy}},\ }\href
  {https://doi.org/https://doi.org/10.1016/0370-2693(80)90825-4} {\bibfield
  {journal} {\bibinfo  {journal} {Phys. Lett.}\ }\textbf {\bibinfo {volume}
  {B94}},\ \bibinfo {pages} {61 } (\bibinfo {year} {1980})}\BibitemShut
  {NoStop}%
\bibitem [{\citenamefont {Cheng}\ and\ \citenamefont {Li}(1980)}]{seesawII2}%
  \BibitemOpen
  \bibfield  {author} {\bibinfo {author} {\bibfnamefont {T.~P.}\ \bibnamefont
  {Cheng}}\ and\ \bibinfo {author} {\bibfnamefont {L.-F.}\ \bibnamefont {Li}},\
  }\bibfield  {title} {\bibinfo {title} {{\it Neutrino masses, mixings, and
  oscillations in SU(2)$\times$U(1) models of electroweak interactions}},\
  }\href {https://doi.org/10.1103/PhysRevD.22.2860} {\bibfield  {journal}
  {\bibinfo  {journal} {Phys. Rev.}\ }\textbf {\bibinfo {volume} {D22}},\
  \bibinfo {pages} {2860} (\bibinfo {year} {1980})}\BibitemShut {NoStop}%
\bibitem [{\citenamefont {Schechter}\ and\ \citenamefont
  {Valle}(1980)}]{seesawII3}%
  \BibitemOpen
  \bibfield  {author} {\bibinfo {author} {\bibfnamefont {J.}~\bibnamefont
  {Schechter}}\ and\ \bibinfo {author} {\bibfnamefont {J.~W.~F.}\ \bibnamefont
  {Valle}},\ }\bibfield  {title} {\bibinfo {title} {{\it Neutrino masses in
  SU(2)$\times$U(1) theories}},\ }\href
  {https://doi.org/10.1103/PhysRevD.22.2227} {\bibfield  {journal} {\bibinfo
  {journal} {Phys. Rev.}\ }\textbf {\bibinfo {volume} {D22}},\ \bibinfo {pages}
  {2227} (\bibinfo {year} {1980})}\BibitemShut {NoStop}%
\bibitem [{\citenamefont {Lazarides}\ \emph {et~al.}(1981)\citenamefont
  {Lazarides}, \citenamefont {Shafi},\ and\ \citenamefont
  {Wetterich}}]{seesawII4}%
  \BibitemOpen
  \bibfield  {author} {\bibinfo {author} {\bibfnamefont {G.}~\bibnamefont
  {Lazarides}}, \bibinfo {author} {\bibfnamefont {Q.}~\bibnamefont {Shafi}},\
  and\ \bibinfo {author} {\bibfnamefont {C.}~\bibnamefont {Wetterich}},\
  }\bibfield  {title} {\bibinfo {title} {{\it Proton lifetime and fermion
  masses in an SO(10) model}},\ }\href
  {https://doi.org/https://doi.org/10.1016/0550-3213(81)90354-0} {\bibfield
  {journal} {\bibinfo  {journal} {Nucl. Phys.}\ }\textbf {\bibinfo {volume}
  {B181}},\ \bibinfo {pages} {287 } (\bibinfo {year} {1981})}\BibitemShut
  {NoStop}%
\bibitem [{\citenamefont {Mohapatra}\ and\ \citenamefont
  {Senjanovi\ifmmode~\acute{c}\else \'{c}\fi{}}(1981)}]{seesawII5}%
  \BibitemOpen
  \bibfield  {author} {\bibinfo {author} {\bibfnamefont {R.~N.}\ \bibnamefont
  {Mohapatra}}\ and\ \bibinfo {author} {\bibfnamefont {G.}~\bibnamefont
  {Senjanovi\ifmmode~\acute{c}\else \'{c}\fi{}}},\ }\bibfield  {title}
  {\bibinfo {title} {{\it Neutrino masses and mixings in gauge models with
  spontaneous parity violation}},\ }\href
  {https://doi.org/10.1103/PhysRevD.23.165} {\bibfield  {journal} {\bibinfo
  {journal} {Phys. Rev.}\ }\textbf {\bibinfo {volume} {D23}},\ \bibinfo {pages}
  {165} (\bibinfo {year} {1981})}\BibitemShut {NoStop}%
\bibitem [{\citenamefont {Aaboud}\ \emph
  {et~al.}(2018{\natexlab{c}})\citenamefont {Aaboud} \emph
  {et~al.}}]{resonanceWZ}%
  \BibitemOpen
  \bibfield  {author} {\bibinfo {author} {\bibfnamefont {M.}~\bibnamefont
  {Aaboud}} \emph {et~al.} (\bibinfo {collaboration} {ATLAS}),\ }\bibfield
  {title} {\bibinfo {title} {{\it Search for resonant $WZ$ production in the
  fully leptonic final state in proton-proton collisions at $\sqrt{s} = 13$ TeV
  with the ATLAS detector}},\ }\href
  {https://doi.org/10.1016/j.physletb.2018.10.021} {\bibfield  {journal}
  {\bibinfo  {journal} {Phys. Lett.}\ }\textbf {\bibinfo {volume} {B787}},\
  \bibinfo {pages} {68} (\bibinfo {year} {2018}{\natexlab{c}})},\ \Eprint
  {https://arxiv.org/abs/1806.01532} {arXiv:1806.01532 [hep-ex]} \BibitemShut
  {NoStop}%
%%CITATION = ARXIV:1806.01532;%%
\bibitem [{\citenamefont {Aad}\ \emph {et~al.}(2015)\citenamefont {Aad} \emph
  {et~al.}}]{WZ}%
  \BibitemOpen
  \bibfield  {author} {\bibinfo {author} {\bibfnamefont {G.}~\bibnamefont
  {Aad}} \emph {et~al.} (\bibinfo {collaboration} {ATLAS}),\ }\bibfield
  {title} {\bibinfo {title} {{\it Search for a Charged Higgs Boson Produced in
  the Vector-Boson Fusion Mode with Decay $H^\pm \to W^\pm Z$ using $pp$
  Collisions at $\sqrt{s}=8$ TeV with the ATLAS Experiment}},\ }\href
  {https://doi.org/10.1103/PhysRevLett.114.231801} {\bibfield  {journal}
  {\bibinfo  {journal} {Phys. Rev. Lett.}\ }\textbf {\bibinfo {volume} {114}},\
  \bibinfo {pages} {231801} (\bibinfo {year} {2015})},\ \Eprint
  {https://arxiv.org/abs/1503.04233} {arXiv:1503.04233 [hep-ex]} \BibitemShut
  {NoStop}%
%%CITATION = ARXIV:1503.04233;%%
\bibitem [{\citenamefont {Aaboud}\ \emph {et~al.}(2019)\citenamefont {Aaboud}
  \emph {et~al.}}]{WW}%
  \BibitemOpen
  \bibfield  {author} {\bibinfo {author} {\bibfnamefont {M.}~\bibnamefont
  {Aaboud}} \emph {et~al.} (\bibinfo {collaboration} {ATLAS}),\ }\bibfield
  {title} {\bibinfo {title} {{\it Search for doubly charged scalar bosons
  decaying into same-sign $W$ boson pairs with the ATLAS detector}},\ }\href
  {https://doi.org/10.1140/epjc/s10052-018-6500-y} {\bibfield  {journal}
  {\bibinfo  {journal} {Eur. Phys. J.}\ }\textbf {\bibinfo {volume} {C79}},\
  \bibinfo {pages} {58} (\bibinfo {year} {2019})},\ \Eprint
  {https://arxiv.org/abs/1808.01899} {arXiv:1808.01899 [hep-ex]} \BibitemShut
  {NoStop}%
%%CITATION = ARXIV:1808.01899;%%
\bibitem [{\citenamefont {Zee}(1980)}]{ref:Zee_model}%
  \BibitemOpen
  \bibfield  {author} {\bibinfo {author} {\bibfnamefont {A.}~\bibnamefont
  {Zee}},\ }\bibfield  {title} {\bibinfo {title} {{\it A Theory of Lepton
  Number Violation, Neutrino Majorana Mass, and Oscillation}},\ }\href
  {https://doi.org/10.1016/0370-2693(80)90349-4, 10.1016/0370-2693(80)90193-8}
  {\bibfield  {journal} {\bibinfo  {journal} {Phys. Lett.}\ }\textbf {\bibinfo
  {volume} {B93}},\ \bibinfo {pages} {389} (\bibinfo {year} {1980})},\ \bibinfo
  {note} {[Erratum: Phys. Lett.95B,461(1980)]}\BibitemShut {NoStop}%
%%CITATION = PHLTA,93B,389;%%
\bibitem [{\citenamefont {Zee}(1986)}]{ref:Zee-Babu_model1}%
  \BibitemOpen
  \bibfield  {author} {\bibinfo {author} {\bibfnamefont {A.}~\bibnamefont
  {Zee}},\ }\bibfield  {title} {\bibinfo {title} {{\it Quantum Numbers of
  Majorana Neutrino Masses}},\ }\href
  {https://doi.org/10.1016/0550-3213(86)90475-X} {\bibfield  {journal}
  {\bibinfo  {journal} {Nucl. Phys.}\ }\textbf {\bibinfo {volume} {B264}},\
  \bibinfo {pages} {99} (\bibinfo {year} {1986})}\BibitemShut {NoStop}%
%%CITATION = NUPHA,B264,99;%%
\bibitem [{\citenamefont {Babu}(1988)}]{ref:Zee-Babu_model2}%
  \BibitemOpen
  \bibfield  {author} {\bibinfo {author} {\bibfnamefont {K.~S.}\ \bibnamefont
  {Babu}},\ }\bibfield  {title} {\bibinfo {title} {{\it Model of 'Calculable'
  Majorana Neutrino Masses}},\ }\href
  {https://doi.org/10.1016/0370-2693(88)91584-5} {\bibfield  {journal}
  {\bibinfo  {journal} {Phys. Lett.}\ }\textbf {\bibinfo {volume} {B203}},\
  \bibinfo {pages} {132} (\bibinfo {year} {1988})}\BibitemShut {NoStop}%
%%CITATION = PHLTA,B203,132;%%
\bibitem [{\citenamefont {Martin}(1997)}]{Martin}%
  \BibitemOpen
  \bibfield  {author} {\bibinfo {author} {\bibfnamefont {S.~P.}\ \bibnamefont
  {Martin}},\ }\bibfield  {title} {\bibinfo {title} {{\it A Supersymmetry
  primer}},\ }\href {https://doi.org/10.1142/9789812839657_0001,
  10.1142/9789814307505_0001} {\bibfield  {journal} {\bibinfo  {journal} {[Adv.
  Ser. Direct. High Energy Phys.18,1(1998)]}\ ,\ \bibinfo {pages} {1}}
  (\bibinfo {year} {1997})},\ \Eprint {https://arxiv.org/abs/hep-ph/9709356}
  {arXiv:hep-ph/9709356 [hep-ph]} \BibitemShut {NoStop}%
%%CITATION = HEP-PH/9709356;%%
\bibitem [{\citenamefont {Fileviez~Perez}\ and\ \citenamefont
  {Murgui}(2017)}]{ref:L-Rmodel}%
  \BibitemOpen
  \bibfield  {author} {\bibinfo {author} {\bibfnamefont {P.}~\bibnamefont
  {Fileviez~Perez}}\ and\ \bibinfo {author} {\bibfnamefont {C.}~\bibnamefont
  {Murgui}},\ }\bibfield  {title} {\bibinfo {title} {{\it Lepton Flavour
  Violation in Left-Right Theory}},\ }\href
  {https://doi.org/10.1103/PhysRevD.95.075010} {\bibfield  {journal} {\bibinfo
  {journal} {Phys. Rev.}\ }\textbf {\bibinfo {volume} {D95}},\ \bibinfo {pages}
  {075010} (\bibinfo {year} {2017})},\ \Eprint
  {https://arxiv.org/abs/1701.06801} {arXiv:1701.06801 [hep-ph]} \BibitemShut
  {NoStop}%
%%CITATION = ARXIV:1701.06801;%%
\bibitem [{\citenamefont {Cao}\ \emph {et~al.}(2018)\citenamefont {Cao},
  \citenamefont {Li}, \citenamefont {Xie},\ and\ \citenamefont
  {Zhang}}]{Cao:2017ffm}%
  \BibitemOpen
  \bibfield  {author} {\bibinfo {author} {\bibfnamefont {Q.-H.}\ \bibnamefont
  {Cao}}, \bibinfo {author} {\bibfnamefont {G.}~\bibnamefont {Li}}, \bibinfo
  {author} {\bibfnamefont {K.-P.}\ \bibnamefont {Xie}},\ and\ \bibinfo {author}
  {\bibfnamefont {J.}~\bibnamefont {Zhang}},\ }\bibfield  {title} {\bibinfo
  {title} {{\it Searching for Weak Singlet Charged Scalar at the Large Hadron
  Collider}},\ }\href {https://doi.org/10.1103/PhysRevD.97.115036} {\bibfield
  {journal} {\bibinfo  {journal} {Phys. Rev.}\ }\textbf {\bibinfo {volume}
  {D97}},\ \bibinfo {pages} {115036} (\bibinfo {year} {2018})},\ \Eprint
  {https://arxiv.org/abs/1711.02113} {arXiv:1711.02113 [hep-ph]} \BibitemShut
  {NoStop}%
%%CITATION = ARXIV:1711.02113;%%
\bibitem [{\citenamefont {Aaboud}\ \emph
  {et~al.}(2018{\natexlab{d}})\citenamefont {Aaboud} \emph
  {et~al.}}]{dilepton2018}%
  \BibitemOpen
  \bibfield  {author} {\bibinfo {author} {\bibfnamefont {M.}~\bibnamefont
  {Aaboud}} \emph {et~al.} (\bibinfo {collaboration} {ATLAS}),\ }\bibfield
  {title} {\bibinfo {title} {{\it Search for electroweak production of
  supersymmetric particles in final states with two or three leptons at
  $\sqrt{s}=13\,$TeV with the ATLAS detector}},\ }\href
  {https://doi.org/10.1140/epjc/s10052-018-6423-7} {\bibfield  {journal}
  {\bibinfo  {journal} {Eur. Phys. J.}\ }\textbf {\bibinfo {volume} {C78}},\
  \bibinfo {pages} {995} (\bibinfo {year} {2018}{\natexlab{d}})},\ \Eprint
  {https://arxiv.org/abs/1803.02762} {arXiv:1803.02762 [hep-ex]} \BibitemShut
  {NoStop}%
%%CITATION = ARXIV:1803.02762;%%
\bibitem [{\citenamefont {{ATLAS Collaboration}}(2019)}]{ditau2019}%
  \BibitemOpen
  \bibfield  {author} {\bibinfo {author} {\bibnamefont {{ATLAS
  Collaboration}}},\ }\href {https://cds.cern.ch/record/2676595} {\emph
  {\bibinfo {title} {{\it Search for direct stau production in events with two
  hadronic tau leptons in √s = 13 TeV pp collisions with the ATLAS
  detector}}}},\ \bibinfo {type} {Tech. Rep.}\ \bibinfo {number}
  {ATLAS-CONF-2019-018}\ (\bibinfo  {institution} {CERN},\ \bibinfo {address}
  {Geneva},\ \bibinfo {year} {2019})\BibitemShut {NoStop}%
\bibitem [{\citenamefont {McLaughlin}\ and\ \citenamefont
  {Ng}(1999)}]{McLaughlin:1999rr}%
  \BibitemOpen
  \bibfield  {author} {\bibinfo {author} {\bibfnamefont {G.~C.}\ \bibnamefont
  {McLaughlin}}\ and\ \bibinfo {author} {\bibfnamefont {J.~N.}\ \bibnamefont
  {Ng}},\ }\bibfield  {title} {\bibinfo {title} {{{\it A Study of the charged
  scalar in the Zee model}}},\ }\href
  {https://doi.org/10.1016/S0370-2693(99)00490-6} {\bibfield  {journal}
  {\bibinfo  {journal} {Phys. Lett.}\ }\textbf {\bibinfo {volume} {B455}},\
  \bibinfo {pages} {224} (\bibinfo {year} {1999})},\ \Eprint
  {https://arxiv.org/abs/hep-ph/9903509} {arXiv:hep-ph/9903509 [hep-ph]}
  \BibitemShut {NoStop}%
%%CITATION = HEP-PH/9903509;%%
\bibitem [{\citenamefont {Herrero-Garcia}\ \emph {et~al.}(2014)\citenamefont
  {Herrero-Garcia}, \citenamefont {Nebot}, \citenamefont {Rius},\ and\
  \citenamefont {Santamaria}}]{Herrero-Garcia:2014hfa}%
  \BibitemOpen
  \bibfield  {author} {\bibinfo {author} {\bibfnamefont {J.}~\bibnamefont
  {Herrero-Garcia}}, \bibinfo {author} {\bibfnamefont {M.}~\bibnamefont
  {Nebot}}, \bibinfo {author} {\bibfnamefont {N.}~\bibnamefont {Rius}},\ and\
  \bibinfo {author} {\bibfnamefont {A.}~\bibnamefont {Santamaria}},\ }\bibfield
   {title} {\bibinfo {title} {{\it The Zee--Babu model revisited in the light
  of new data}},\ }\href {https://doi.org/10.1016/j.nuclphysb.2014.06.001}
  {\bibfield  {journal} {\bibinfo  {journal} {Nucl. Phys.}\ }\textbf {\bibinfo
  {volume} {B885}},\ \bibinfo {pages} {542} (\bibinfo {year} {2014})},\ \Eprint
  {https://arxiv.org/abs/1402.4491} {arXiv:1402.4491 [hep-ph]} \BibitemShut
  {NoStop}%
%%CITATION = ARXIV:1402.4491;%%
\bibitem [{\citenamefont {Baldini}\ \emph {et~al.}(2016)\citenamefont {Baldini}
  \emph {et~al.}}]{TheMEG:2016wtm}%
  \BibitemOpen
  \bibfield  {author} {\bibinfo {author} {\bibfnamefont {A.~M.}\ \bibnamefont
  {Baldini}} \emph {et~al.} (\bibinfo {collaboration} {MEG}),\ }\bibfield
  {title} {\bibinfo {title} {{{\it Search for the lepton flavour violating
  decay $\mu ^+ \rightarrow \mathrm {e}^+ \gamma $ with the full dataset of the
  MEG experiment}}},\ }\href {https://doi.org/10.1140/epjc/s10052-016-4271-x}
  {\bibfield  {journal} {\bibinfo  {journal} {Eur. Phys. J.}\ }\textbf
  {\bibinfo {volume} {C76}},\ \bibinfo {pages} {434} (\bibinfo {year}
  {2016})},\ \Eprint {https://arxiv.org/abs/1605.05081} {arXiv:1605.05081
  [hep-ex]} \BibitemShut {NoStop}%
%%CITATION = ARXIV:1605.05081;%%
\bibitem [{\citenamefont {Tanabashi}\ \emph {et~al.}(2018)\citenamefont
  {Tanabashi} \emph {et~al.}}]{Tanabashi:2018oca}%
  \BibitemOpen
  \bibfield  {author} {\bibinfo {author} {\bibfnamefont {M.}~\bibnamefont
  {Tanabashi}} \emph {et~al.} (\bibinfo {collaboration} {Particle Data
  Group}),\ }\bibfield  {title} {\bibinfo {title} {{\it Review of Particle
  Physics}},\ }\href {https://doi.org/10.1103/PhysRevD.98.030001} {\bibfield
  {journal} {\bibinfo  {journal} {Phys. Rev.}\ }\textbf {\bibinfo {volume}
  {D98}},\ \bibinfo {pages} {030001} (\bibinfo {year} {2018})}\BibitemShut
  {NoStop}%
%%CITATION = PHRVA,D98,030001;%%
\bibitem [{\citenamefont {Alwall}\ \emph {et~al.}(2014)\citenamefont {Alwall},
  \citenamefont {Frederix}, \citenamefont {Frixione}, \citenamefont {Hirschi},
  \citenamefont {Maltoni}, \citenamefont {Mattelaer}, \citenamefont {Shao},
  \citenamefont {Stelzer}, \citenamefont {Torrielli},\ and\ \citenamefont
  {Zaro}}]{Alwall:2014hca}%
  \BibitemOpen
  \bibfield  {author} {\bibinfo {author} {\bibfnamefont {J.}~\bibnamefont
  {Alwall}}, \bibinfo {author} {\bibfnamefont {R.}~\bibnamefont {Frederix}},
  \bibinfo {author} {\bibfnamefont {S.}~\bibnamefont {Frixione}}, \bibinfo
  {author} {\bibfnamefont {V.}~\bibnamefont {Hirschi}}, \bibinfo {author}
  {\bibfnamefont {F.}~\bibnamefont {Maltoni}}, \bibinfo {author} {\bibfnamefont
  {O.}~\bibnamefont {Mattelaer}}, \bibinfo {author} {\bibfnamefont {H.~S.}\
  \bibnamefont {Shao}}, \bibinfo {author} {\bibfnamefont {T.}~\bibnamefont
  {Stelzer}}, \bibinfo {author} {\bibfnamefont {P.}~\bibnamefont {Torrielli}},\
  and\ \bibinfo {author} {\bibfnamefont {M.}~\bibnamefont {Zaro}},\ }\bibfield
  {title} {\bibinfo {title} {{\it The automated computation of tree-level and
  next-to-leading order differential cross sections, and their matching to
  parton shower simulations}},\ }\href
  {https://doi.org/10.1007/JHEP07(2014)079} {\bibfield  {journal} {\bibinfo
  {journal} {JHEP}\ }\textbf {\bibinfo {volume} {07}},\ \bibinfo {pages}
  {079}},\ \Eprint {https://arxiv.org/abs/1405.0301} {arXiv:1405.0301 [hep-ph]}
  \BibitemShut {NoStop}%
%%CITATION = ARXIV:1405.0301;%%
\bibitem [{\citenamefont {Sj{\"o}strand}\ \emph {et~al.}(2015)\citenamefont
  {Sj{\"o}strand}, \citenamefont {Ask}, \citenamefont {Christiansen},
  \citenamefont {Corke}, \citenamefont {Desai}, \citenamefont {Ilten},
  \citenamefont {Mrenna}, \citenamefont {Prestel}, \citenamefont {Rasmussen},\
  and\ \citenamefont {Skands}}]{Sjostrand:2014zea}%
  \BibitemOpen
  \bibfield  {author} {\bibinfo {author} {\bibfnamefont {T.}~\bibnamefont
  {Sj{\"o}strand}}, \bibinfo {author} {\bibfnamefont {S.}~\bibnamefont {Ask}},
  \bibinfo {author} {\bibfnamefont {J.~R.}\ \bibnamefont {Christiansen}},
  \bibinfo {author} {\bibfnamefont {R.}~\bibnamefont {Corke}}, \bibinfo
  {author} {\bibfnamefont {N.}~\bibnamefont {Desai}}, \bibinfo {author}
  {\bibfnamefont {P.}~\bibnamefont {Ilten}}, \bibinfo {author} {\bibfnamefont
  {S.}~\bibnamefont {Mrenna}}, \bibinfo {author} {\bibfnamefont
  {S.}~\bibnamefont {Prestel}}, \bibinfo {author} {\bibfnamefont {C.~O.}\
  \bibnamefont {Rasmussen}},\ and\ \bibinfo {author} {\bibfnamefont {P.~Z.}\
  \bibnamefont {Skands}},\ }\bibfield  {title} {\bibinfo {title} {{\it An
  Introduction to PYTHIA 8.2}},\ }\href
  {https://doi.org/10.1016/j.cpc.2015.01.024} {\bibfield  {journal} {\bibinfo
  {journal} {Comput. Phys. Commun.}\ }\textbf {\bibinfo {volume} {191}},\
  \bibinfo {pages} {159} (\bibinfo {year} {2015})},\ \Eprint
  {https://arxiv.org/abs/1410.3012} {arXiv:1410.3012 [hep-ph]} \BibitemShut
  {NoStop}%
%%CITATION = ARXIV:1410.3012;%%
\bibitem [{\citenamefont {de~Favereau}\ \emph {et~al.}(2014)\citenamefont
  {de~Favereau}, \citenamefont {Delaere}, \citenamefont {Demin}, \citenamefont
  {Giammanco}, \citenamefont {Lema{\^\i}tre}, \citenamefont {Mertens},\ and\
  \citenamefont {Selvaggi}}]{deFavereau:2013fsa}%
  \BibitemOpen
  \bibfield  {author} {\bibinfo {author} {\bibfnamefont {J.}~\bibnamefont
  {de~Favereau}}, \bibinfo {author} {\bibfnamefont {C.}~\bibnamefont
  {Delaere}}, \bibinfo {author} {\bibfnamefont {P.}~\bibnamefont {Demin}},
  \bibinfo {author} {\bibfnamefont {A.}~\bibnamefont {Giammanco}}, \bibinfo
  {author} {\bibfnamefont {V.}~\bibnamefont {Lema{\^\i}tre}}, \bibinfo {author}
  {\bibfnamefont {A.}~\bibnamefont {Mertens}},\ and\ \bibinfo {author}
  {\bibfnamefont {M.}~\bibnamefont {Selvaggi}} (\bibinfo {collaboration}
  {DELPHES 3}),\ }\bibfield  {title} {\bibinfo {title} {{\it DELPHES 3, A
  modular framework for fast simulation of a generic collider experiment}},\
  }\href {https://doi.org/10.1007/JHEP02(2014)057} {\bibfield  {journal}
  {\bibinfo  {journal} {JHEP}\ }\textbf {\bibinfo {volume} {02}},\ \bibinfo
  {pages} {057}},\ \Eprint {https://arxiv.org/abs/1307.6346} {arXiv:1307.6346
  [hep-ex]} \BibitemShut {NoStop}%
%%CITATION = ARXIV:1307.6346;%%
\bibitem [{\citenamefont {Alloul}\ \emph {et~al.}(2014)\citenamefont {Alloul},
  \citenamefont {Christensen}, \citenamefont {Degrande}, \citenamefont {Duhr},\
  and\ \citenamefont {Fuks}}]{Alloul:2013bka}%
  \BibitemOpen
  \bibfield  {author} {\bibinfo {author} {\bibfnamefont {A.}~\bibnamefont
  {Alloul}}, \bibinfo {author} {\bibfnamefont {N.~D.}\ \bibnamefont
  {Christensen}}, \bibinfo {author} {\bibfnamefont {C.}~\bibnamefont
  {Degrande}}, \bibinfo {author} {\bibfnamefont {C.}~\bibnamefont {Duhr}},\
  and\ \bibinfo {author} {\bibfnamefont {B.}~\bibnamefont {Fuks}},\ }\bibfield
  {title} {\bibinfo {title} {{\it FeynRules 2.0 - A complete toolbox for
  tree-level phenomenology}},\ }\href
  {https://doi.org/10.1016/j.cpc.2014.04.012} {\bibfield  {journal} {\bibinfo
  {journal} {Comput. Phys. Commun.}\ }\textbf {\bibinfo {volume} {185}},\
  \bibinfo {pages} {2250} (\bibinfo {year} {2014})},\ \Eprint
  {https://arxiv.org/abs/1310.1921} {arXiv:1310.1921 [hep-ph]} \BibitemShut
  {NoStop}%
%%CITATION = ARXIV:1310.1921;%%
\bibitem [{\citenamefont {Degrande}\ \emph {et~al.}(2012)\citenamefont
  {Degrande}, \citenamefont {Duhr}, \citenamefont {Fuks}, \citenamefont
  {Grellscheid}, \citenamefont {Mattelaer},\ and\ \citenamefont
  {Reiter}}]{Degrande:2011ua}%
  \BibitemOpen
  \bibfield  {author} {\bibinfo {author} {\bibfnamefont {C.}~\bibnamefont
  {Degrande}}, \bibinfo {author} {\bibfnamefont {C.}~\bibnamefont {Duhr}},
  \bibinfo {author} {\bibfnamefont {B.}~\bibnamefont {Fuks}}, \bibinfo {author}
  {\bibfnamefont {D.}~\bibnamefont {Grellscheid}}, \bibinfo {author}
  {\bibfnamefont {O.}~\bibnamefont {Mattelaer}},\ and\ \bibinfo {author}
  {\bibfnamefont {T.}~\bibnamefont {Reiter}},\ }\bibfield  {title} {\bibinfo
  {title} {{\it UFO - The Universal FeynRules Output}},\ }\href
  {https://doi.org/10.1016/j.cpc.2012.01.022} {\bibfield  {journal} {\bibinfo
  {journal} {Comput. Phys. Commun.}\ }\textbf {\bibinfo {volume} {183}},\
  \bibinfo {pages} {1201} (\bibinfo {year} {2012})},\ \Eprint
  {https://arxiv.org/abs/1108.2040} {arXiv:1108.2040 [hep-ph]} \BibitemShut
  {NoStop}%
%%CITATION = ARXIV:1108.2040;%%
\bibitem [{\citenamefont {Grazzini}\ \emph
  {et~al.}(2016{\natexlab{a}})\citenamefont {Grazzini}, \citenamefont
  {Kallweit}, \citenamefont {Pozzorini}, \citenamefont {Rathlev},\ and\
  \citenamefont {Wiesemann}}]{Grazzini:2016ctr}%
  \BibitemOpen
  \bibfield  {author} {\bibinfo {author} {\bibfnamefont {M.}~\bibnamefont
  {Grazzini}}, \bibinfo {author} {\bibfnamefont {S.}~\bibnamefont {Kallweit}},
  \bibinfo {author} {\bibfnamefont {S.}~\bibnamefont {Pozzorini}}, \bibinfo
  {author} {\bibfnamefont {D.}~\bibnamefont {Rathlev}},\ and\ \bibinfo {author}
  {\bibfnamefont {M.}~\bibnamefont {Wiesemann}},\ }\bibfield  {title} {\bibinfo
  {title} {{\it $W^+W^-$ production at the LHC: fiducial cross sections and
  distributions in NNLO QCD}},\ }\href
  {https://doi.org/10.1007/JHEP08(2016)140} {\bibfield  {journal} {\bibinfo
  {journal} {JHEP}\ }\textbf {\bibinfo {volume} {08}},\ \bibinfo {pages}
  {140}},\ \Eprint {https://arxiv.org/abs/1605.02716} {arXiv:1605.02716
  [hep-ph]} \BibitemShut {NoStop}%
%%CITATION = ARXIV:1605.02716;%%
\bibitem [{\citenamefont {Caola}\ \emph {et~al.}(2016)\citenamefont {Caola},
  \citenamefont {Melnikov}, \citenamefont {R{\"o}ntsch},\ and\ \citenamefont
  {Tancredi}}]{Caola:2015rqy}%
  \BibitemOpen
  \bibfield  {author} {\bibinfo {author} {\bibfnamefont {F.}~\bibnamefont
  {Caola}}, \bibinfo {author} {\bibfnamefont {K.}~\bibnamefont {Melnikov}},
  \bibinfo {author} {\bibfnamefont {R.}~\bibnamefont {R{\"o}ntsch}},\ and\
  \bibinfo {author} {\bibfnamefont {L.}~\bibnamefont {Tancredi}},\ }\bibfield
  {title} {\bibinfo {title} {{\it QCD corrections to $W^+W^-$ production
  through gluon fusion}},\ }\href
  {https://doi.org/10.1016/j.physletb.2016.01.046} {\bibfield  {journal}
  {\bibinfo  {journal} {Phys. Lett.}\ }\textbf {\bibinfo {volume} {B754}},\
  \bibinfo {pages} {275} (\bibinfo {year} {2016})},\ \Eprint
  {https://arxiv.org/abs/1511.08617} {arXiv:1511.08617 [hep-ph]} \BibitemShut
  {NoStop}%
%%CITATION = ARXIV:1511.08617;%%
\bibitem [{\citenamefont {Aad}\ \emph {et~al.}(2016)\citenamefont {Aad} \emph
  {et~al.}}]{Aad:2015zqe}%
  \BibitemOpen
  \bibfield  {author} {\bibinfo {author} {\bibfnamefont {G.}~\bibnamefont
  {Aad}} \emph {et~al.} (\bibinfo {collaboration} {ATLAS}),\ }\bibfield
  {title} {\bibinfo {title} {{\it Measurement of the $ZZ$ Production Cross
  Section in $pp$ Collisions at $\sqrt{s}$ = 13 TeV with the ATLAS Detector}},\
  }\href {https://doi.org/10.1103/PhysRevLett.116.101801} {\bibfield  {journal}
  {\bibinfo  {journal} {Phys. Rev. Lett.}\ }\textbf {\bibinfo {volume} {116}},\
  \bibinfo {pages} {101801} (\bibinfo {year} {2016})},\ \Eprint
  {https://arxiv.org/abs/1512.05314} {arXiv:1512.05314 [hep-ex]} \BibitemShut
  {NoStop}%
%%CITATION = ARXIV:1512.05314;%%
\bibitem [{\citenamefont {Grazzini}\ \emph
  {et~al.}(2016{\natexlab{b}})\citenamefont {Grazzini}, \citenamefont
  {Kallweit}, \citenamefont {Rathlev},\ and\ \citenamefont
  {Wiesemann}}]{Grazzini:2016swo}%
  \BibitemOpen
  \bibfield  {author} {\bibinfo {author} {\bibfnamefont {M.}~\bibnamefont
  {Grazzini}}, \bibinfo {author} {\bibfnamefont {S.}~\bibnamefont {Kallweit}},
  \bibinfo {author} {\bibfnamefont {D.}~\bibnamefont {Rathlev}},\ and\ \bibinfo
  {author} {\bibfnamefont {M.}~\bibnamefont {Wiesemann}},\ }\bibfield  {title}
  {\bibinfo {title} {{\it $W^{\pm}Z$ production at hadron colliders in NNLO
  QCD}},\ }\href {https://doi.org/10.1016/j.physletb.2016.08.017} {\bibfield
  {journal} {\bibinfo  {journal} {Phys. Lett.}\ }\textbf {\bibinfo {volume}
  {B761}},\ \bibinfo {pages} {179} (\bibinfo {year} {2016}{\natexlab{b}})},\
  \Eprint {https://arxiv.org/abs/1604.08576} {arXiv:1604.08576 [hep-ph]}
  \BibitemShut {NoStop}%
%%CITATION = ARXIV:1604.08576;%%
\bibitem [{\citenamefont {Boughezal}\ \emph {et~al.}(2017)\citenamefont
  {Boughezal}, \citenamefont {Campbell}, \citenamefont {Ellis}, \citenamefont
  {Focke}, \citenamefont {Giele}, \citenamefont {Liu}, \citenamefont
  {Petriello},\ and\ \citenamefont {Williams}}]{Boughezal:2016wmq}%
  \BibitemOpen
  \bibfield  {author} {\bibinfo {author} {\bibfnamefont {R.}~\bibnamefont
  {Boughezal}}, \bibinfo {author} {\bibfnamefont {J.~M.}\ \bibnamefont
  {Campbell}}, \bibinfo {author} {\bibfnamefont {R.~K.}\ \bibnamefont {Ellis}},
  \bibinfo {author} {\bibfnamefont {C.}~\bibnamefont {Focke}}, \bibinfo
  {author} {\bibfnamefont {W.}~\bibnamefont {Giele}}, \bibinfo {author}
  {\bibfnamefont {X.}~\bibnamefont {Liu}}, \bibinfo {author} {\bibfnamefont
  {F.}~\bibnamefont {Petriello}},\ and\ \bibinfo {author} {\bibfnamefont
  {C.}~\bibnamefont {Williams}},\ }\bibfield  {title} {\bibinfo {title} {{\it
  Color singlet production at NNLO in MCFM}},\ }\href
  {https://doi.org/10.1140/epjc/s10052-016-4558-y} {\bibfield  {journal}
  {\bibinfo  {journal} {Eur. Phys. J.}\ }\textbf {\bibinfo {volume} {C77}},\
  \bibinfo {pages} {7} (\bibinfo {year} {2017})},\ \Eprint
  {https://arxiv.org/abs/1605.08011} {arXiv:1605.08011 [hep-ph]} \BibitemShut
  {NoStop}%
%%CITATION = ARXIV:1605.08011;%%
\bibitem [{\citenamefont {Ahrens}\ \emph {et~al.}(2011)\citenamefont {Ahrens},
  \citenamefont {Ferroglia}, \citenamefont {Neubert}, \citenamefont {Pecjak},\
  and\ \citenamefont {Yang}}]{Ahrens:2011px}%
  \BibitemOpen
  \bibfield  {author} {\bibinfo {author} {\bibfnamefont {V.}~\bibnamefont
  {Ahrens}}, \bibinfo {author} {\bibfnamefont {A.}~\bibnamefont {Ferroglia}},
  \bibinfo {author} {\bibfnamefont {M.}~\bibnamefont {Neubert}}, \bibinfo
  {author} {\bibfnamefont {B.~D.}\ \bibnamefont {Pecjak}},\ and\ \bibinfo
  {author} {\bibfnamefont {L.~L.}\ \bibnamefont {Yang}},\ }\bibfield  {title}
  {\bibinfo {title} {{\it Precision predictions for the $t\bar{t}$ production
  cross section at hadron colliders}},\ }\href
  {https://doi.org/10.1016/j.physletb.2011.07.058} {\bibfield  {journal}
  {\bibinfo  {journal} {Phys. Lett.}\ }\textbf {\bibinfo {volume} {B703}},\
  \bibinfo {pages} {135} (\bibinfo {year} {2011})},\ \Eprint
  {https://arxiv.org/abs/1105.5824} {arXiv:1105.5824 [hep-ph]} \BibitemShut
  {NoStop}%
%%CITATION = ARXIV:1105.5824;%%
\bibitem [{\citenamefont {Czakon}\ \emph {et~al.}(2013)\citenamefont {Czakon},
  \citenamefont {Fiedler},\ and\ \citenamefont {Mitov}}]{Czakon:2013goa}%
  \BibitemOpen
  \bibfield  {author} {\bibinfo {author} {\bibfnamefont {M.}~\bibnamefont
  {Czakon}}, \bibinfo {author} {\bibfnamefont {P.}~\bibnamefont {Fiedler}},\
  and\ \bibinfo {author} {\bibfnamefont {A.}~\bibnamefont {Mitov}},\ }\bibfield
   {title} {\bibinfo {title} {{\it Total Top-Quark Pair-Production Cross
  Section at Hadron Colliders Through $O(\alpha^4_S)$}},\ }\href
  {https://doi.org/10.1103/PhysRevLett.110.252004} {\bibfield  {journal}
  {\bibinfo  {journal} {Phys. Rev. Lett.}\ }\textbf {\bibinfo {volume} {110}},\
  \bibinfo {pages} {252004} (\bibinfo {year} {2013})},\ \Eprint
  {https://arxiv.org/abs/1303.6254} {arXiv:1303.6254 [hep-ph]} \BibitemShut
  {NoStop}%
%%CITATION = ARXIV:1303.6254;%%
\bibitem [{\citenamefont {Cao}(2008)}]{Cao:2008af}%
  \BibitemOpen
  \bibfield  {author} {\bibinfo {author} {\bibfnamefont {Q.-H.}\ \bibnamefont
  {Cao}},\ }\bibfield  {title} {\bibinfo {title} {{\it Demonstration of One
  Cutoff Phase Space Slicing Method: Next-to-Leading Order QCD Corrections to
  the tW Associated Production in Hadron Collision}},\ }\href@noop {} {\
  (\bibinfo {year} {2008})},\ \Eprint {https://arxiv.org/abs/0801.1539}
  {arXiv:0801.1539 [hep-ph]} \BibitemShut {NoStop}%
%%CITATION = ARXIV:0801.1539;%%
\bibitem [{\citenamefont {Frixione}\ \emph {et~al.}(2008)\citenamefont
  {Frixione}, \citenamefont {Laenen}, \citenamefont {Motylinski}, \citenamefont
  {Webber},\ and\ \citenamefont {White}}]{Frixione:2008yi}%
  \BibitemOpen
  \bibfield  {author} {\bibinfo {author} {\bibfnamefont {S.}~\bibnamefont
  {Frixione}}, \bibinfo {author} {\bibfnamefont {E.}~\bibnamefont {Laenen}},
  \bibinfo {author} {\bibfnamefont {P.}~\bibnamefont {Motylinski}}, \bibinfo
  {author} {\bibfnamefont {B.~R.}\ \bibnamefont {Webber}},\ and\ \bibinfo
  {author} {\bibfnamefont {C.~D.}\ \bibnamefont {White}},\ }\bibfield  {title}
  {\bibinfo {title} {{\it Single-top hadroproduction in association with a W
  boson}},\ }\href {https://doi.org/10.1088/1126-6708/2008/07/029} {\bibfield
  {journal} {\bibinfo  {journal} {JHEP}\ }\textbf {\bibinfo {volume} {07}},\
  \bibinfo {pages} {029}},\ \Eprint {https://arxiv.org/abs/0805.3067}
  {arXiv:0805.3067 [hep-ph]} \BibitemShut {NoStop}%
%%CITATION = ARXIV:0805.3067;%%
\bibitem [{\citenamefont {Kidonakis}(2015)}]{Kidonakis:2015nna}%
  \BibitemOpen
  \bibfield  {author} {\bibinfo {author} {\bibfnamefont {N.}~\bibnamefont
  {Kidonakis}},\ }\bibfield  {title} {\bibinfo {title} {{\it Theoretical
  results for electroweak-boson and single-top production}},\ }\bibfield
  {booktitle} {\emph {\bibinfo {booktitle} {{Proceedings, 23rd International
  Workshop on Deep-Inelastic Scattering and Related Subjects (DIS 2015):
  Dallas, Texas, USA, April 27-May 01, 2015}}},\ }\href
  {https://doi.org/10.22323/1.247.0170} {\bibfield  {journal} {\bibinfo
  {journal} {PoS}\ }\textbf {\bibinfo {volume} {DIS2015}},\ \bibinfo {pages}
  {170} (\bibinfo {year} {2015})},\ \Eprint {https://arxiv.org/abs/1506.04072}
  {arXiv:1506.04072 [hep-ph]} \BibitemShut {NoStop}%
%%CITATION = ARXIV:1506.04072;%%
\bibitem [{\citenamefont {{CMS Collaboration}}(2017)}]{CMS:2017rio}%
  \BibitemOpen
  \bibfield  {author} {\bibinfo {author} {\bibnamefont {{CMS Collaboration}}},\
  }\href {http://cds.cern.ch/record/2273395} {\emph {\bibinfo {title} {{\it
  Search for pair production of tau sleptons in $\sqrt{s}=13~\mathrm{TeV}$ pp
  collisions in the all-hadronic final state}}}},\ \bibinfo {type} {Tech.
  Rep.}\ \bibinfo {number} {{CMS-PAS-SUS-17-003}}\ (\bibinfo {year}
  {2017})\BibitemShut {NoStop}%
%%CITATION = CMS-PAS-SUS-17-003;%%
\bibitem [{\citenamefont {Alcaide}\ \emph {et~al.}(2018)\citenamefont
  {Alcaide}, \citenamefont {Chala},\ and\ \citenamefont
  {Santamaria}}]{Alcaide:2017dcx}%
  \BibitemOpen
  \bibfield  {author} {\bibinfo {author} {\bibfnamefont {J.}~\bibnamefont
  {Alcaide}}, \bibinfo {author} {\bibfnamefont {M.}~\bibnamefont {Chala}},\
  and\ \bibinfo {author} {\bibfnamefont {A.}~\bibnamefont {Santamaria}},\
  }\bibfield  {title} {\bibinfo {title} {{\it LHC signals of
  radiatively-induced neutrino masses and implications for the Zee--Babu
  model}},\ }\href {https://doi.org/10.1016/j.physletb.2018.02.001} {\bibfield
  {journal} {\bibinfo  {journal} {Phys. Lett.}\ }\textbf {\bibinfo {volume}
  {B779}},\ \bibinfo {pages} {107} (\bibinfo {year} {2018})},\ \Eprint
  {https://arxiv.org/abs/1710.05885} {arXiv:1710.05885 [hep-ph]} \BibitemShut
  {NoStop}%
%%CITATION = ARXIV:1710.05885;%%
\bibitem [{\citenamefont {Sjostrand}\ \emph {et~al.}(2006)\citenamefont
  {Sjostrand}, \citenamefont {Mrenna},\ and\ \citenamefont
  {Skands}}]{ref_pythia6}%
  \BibitemOpen
  \bibfield  {author} {\bibinfo {author} {\bibfnamefont {T.}~\bibnamefont
  {Sjostrand}}, \bibinfo {author} {\bibfnamefont {S.}~\bibnamefont {Mrenna}},\
  and\ \bibinfo {author} {\bibfnamefont {P.~Z.}\ \bibnamefont {Skands}},\
  }\bibfield  {title} {\bibinfo {title} {{\it PYTHIA 6.4 Physics and Manual}},\
  }\href {https://doi.org/10.1088/1126-6708/2006/05/026} {\bibfield  {journal}
  {\bibinfo  {journal} {JHEP}\ }\textbf {\bibinfo {volume} {05}},\ \bibinfo
  {pages} {026}},\ \Eprint {https://arxiv.org/abs/hep-ph/0603175}
  {arXiv:hep-ph/0603175 [hep-ph]} \BibitemShut {NoStop}%
%%CITATION = HEP-PH/0603175;%%
\bibitem [{\citenamefont {Binosi}\ and\ \citenamefont
  {Theussl}(2004)}]{ref:jaxodraw1}%
  \BibitemOpen
  \bibfield  {author} {\bibinfo {author} {\bibfnamefont {D.}~\bibnamefont
  {Binosi}}\ and\ \bibinfo {author} {\bibfnamefont {L.}~\bibnamefont
  {Theussl}},\ }\bibfield  {title} {\bibinfo {title} {{\it JaxoDraw: A
  Graphical user interface for drawing Feynman diagrams}},\ }\href
  {https://doi.org/10.1016/j.cpc.2004.05.001} {\bibfield  {journal} {\bibinfo
  {journal} {Comput. Phys. Commun.}\ }\textbf {\bibinfo {volume} {161}},\
  \bibinfo {pages} {76} (\bibinfo {year} {2004})},\ \Eprint
  {https://arxiv.org/abs/hep-ph/0309015} {arXiv:hep-ph/0309015 [hep-ph]}
  \BibitemShut {NoStop}%
%%CITATION = HEP-PH/0309015;%%
\bibitem [{\citenamefont {Binosi}\ \emph {et~al.}(2009)\citenamefont {Binosi},
  \citenamefont {Collins}, \citenamefont {Kaufhold},\ and\ \citenamefont
  {Theussl}}]{ref:jaxodraw2}%
  \BibitemOpen
  \bibfield  {author} {\bibinfo {author} {\bibfnamefont {D.}~\bibnamefont
  {Binosi}}, \bibinfo {author} {\bibfnamefont {J.}~\bibnamefont {Collins}},
  \bibinfo {author} {\bibfnamefont {C.}~\bibnamefont {Kaufhold}},\ and\
  \bibinfo {author} {\bibfnamefont {L.}~\bibnamefont {Theussl}},\ }\bibfield
  {title} {\bibinfo {title} {{\it JaxoDraw: A Graphical user interface for
  drawing Feynman diagrams. Version 2.0 release notes}},\ }\href
  {https://doi.org/10.1016/j.cpc.2009.02.020} {\bibfield  {journal} {\bibinfo
  {journal} {Comput. Phys. Commun.}\ }\textbf {\bibinfo {volume} {180}},\
  \bibinfo {pages} {1709} (\bibinfo {year} {2009})},\ \Eprint
  {https://arxiv.org/abs/0811.4113} {arXiv:0811.4113 [hep-ph]} \BibitemShut
  {NoStop}%
%%CITATION = ARXIV:0811.4113;%%
\end{thebibliography}%
\end{document}